\documentclass[aps,pre,showpacs]{revtex4}
\usepackage{amsmath}
\usepackage{graphicx}
\usepackage{color}
\usepackage{colordvi}

\begin{document}

\title{Nonlinear management of the miscibility-immiscibility transition in
binary \\
Bose-Einstein condensates}
\author{B. B. Baizakov$^{1,2},$ B. A. Malomed$^{3,4}$, and M. Salerno$^{5}$}
\affiliation{$^{1}$ Physical-Technical Institute, Uzbek Academy of Sciences, 100084,
Tashkent, Uzbekistan,\\
$^{2}$ Institute of Theoretical Physics, National University of Uzbekistan,
Tashkent 100174, Uzbekistan. \\
$^{3}$ Department of Physical Electronics, School of Electrical Engineering,
Faculty of Engineering, Tel Aviv University, Tel Aviv, 69978, Israel.\\
$^{4}$ Instituto de Alta Investigaci\'{o}n, Universidad de Tarapac\'{a},
Casilla 7D, Arica, Chile\\
$^{5}$ Dipartimento di Fisica \textquotedblleft E.R. Caianiello",
Universit\'{a} di Salerno, Via Giovanni Paolo II, 84084 Fisciano
(SA), Italy,}

\begin{abstract}
We investigate application of the nonlinearity management (NM, i.e.,
periodic variation of the strength of the inter-component repulsion)
to the miscibility-immiscibility (MIM) transition across the
critical point of a two-component BEC, both with and without the
linear mixing (Rabi coupling, RC) between the components. To this
end, we first identify, by means of a variational approximation and
numerical solution, diverse stationary domain-wall (DW) structures
supported by the system in the absence of the management. The
approximate analytical solutions for the DWs are found to be in
excellent agreement with their numerical counterparts. An analytical
estimate is also produced for the upshift of the MIM transition
caused by the pressure of the trapping potential in the case of a
confined system. An exact DW solution is produced for the system
including the P\"{o}schl-Teller potential, which is stable
(unstable) if the potential is repulsive (attractive). Further, we
find the spectrum of linear excitations in the spatially uniform
mixed state, and thus establish parameter regions where the system
is stable/unstable against demixing. In particular, RC upshifts the
critical strength of the inter-component repulsion for the onset of
the MIM transition. Eigenfrequencies of excitations on top of DW
states are identified from numerical simulations through monitoring
the evolution of perturbed states. Weak NM applied at the DW
eigenfrequency reveals features of the nonlinear resonance. Stronger
NM, under which the system periodically crosses the MIM-transition
point, restricts the miscibility.
\end{abstract}
\maketitle

\section{Introduction}

Two-component Bose-Einstein condensates (BECs) offer a platform for
the realization of various static states and dynamical effects, in
the form of matter waves. An adequate model of the two-component
mixture is provided by the system of coupled Gross-Pitaevskii (GP)
equations. In particular, a fundamentally important topic is the
miscibility-immiscibility (MIM) transition in the binary condensate
\cite{Mineev,Timmermans}. In this case,
the relation between scattering lengths of inter-atomic collisions, $%
a_{11,22}$, which determine the effective strengths of the
self-repulsion in each component, and $a_{12}$, which determines the
inter-component repulsion strength, plays a crucial role. In the
spatially uniform binary condensate, the mixed state is unstable in
the case of
\begin{equation}
\sqrt{a_{11}a_{22}}<a_{12},  \label{aaa}
\end{equation}
and is stable in the opposite case, while the demixed states, with
spatially separated components, is stable in the former case, and
unstable in the latter one \cite{Mineev,pethick-book}. On the other
hand, if a binary condensate is loaded in a potential trap, the
pressure of the trapping potential naturally makes the BEC more
miscible, upshifting the critical
(demixing) value of $a_{12}$ to one exceeding $\sqrt{a_{11}a_{22}}$ \cite%
{merhasin2005}. An example of this effect is briefly presented
below, for the symmetric case, with $a_{11}=a_{22}\equiv a$: under
the action of the trapping parabolic potential, it follows from Eq.
(\ref{deltaE}) that the instability region (\ref{aaa}) of the mixed
states upshifts from $a<a_{12}$ to $a<(3/2)a_{12}$.

Using the Feshbach-resonance technique \cite{FR}, binary condensates
have been created with tunable inter-species interactions
\cite{Inguscio}, which made it possible to demonstrate MIM phase
transitions \cite{wang2016} and
other effects, such as the creation of quantum droplets \cite%
{QD-ICFO,QD-Florence,QD-Luca} and two-dimensional Townes solitons \cite%
{Bakkali,Bakkali2}. A generic manifestation of the transition to
immiscibility in two-component BECs is the formation of domain walls
(DWs), i.e., boundaries separating immiscible components. The DW\
width is primarily determined by the strength of the inter-component
repulsion. In terms of the underlying system of GP equations, DWs
are represented by compound solutions which include a kink in one
component and an antikink in the other (see, e.g., an exact DW
solution given below by Eqs. (\ref{exact1}) and (\ref{exact2}).

Apparently similar stationary DW patterns, alias \textit{grain
boundaries}, although maintained by different physical mechanisms,
are commonly known in various forms in solid-state physics
\cite{grain1}-\cite{grain3} and in magnetics \cite{grain5,grain6}.
Another realization of DW patterns is well known in the form of
stationary quasi-one-dimensional boundaries between two-dimensional
domains formed by mismatched phases that have a dynamical origin.
These may be DWs produced by collision of propagating fronts in the
framework of the Kibble-Zurek mechanism \cite{KZ,KZ-Laroze}, or
boundaries between Rayleigh-B\'{e}nard-convection patterns of the
roll type, with different orientations
\cite{Cross1982}-\cite{Iooss}.

The formation of DWs in binary BECs, in connection to the MIM
transitions in
these systems, has been explored too, in diverse contexts \cite%
{merhasin2005,filatrella2014,nicklas2011,malomed2004}. In addition
to the empirical analysis reported in those works, a mathematically
rigorous proof of the existence and stability of DWs, as stationary
solutions of the respective system of nonlinearly-coupled GP
equations, was elaborated too, including the system involving a weak
localized potential acting on the binary condensate \cite{Peli}.
These findings suggest that there remain significant subjects for
further studies of the MIM transitions and DWs in
quasi-one-dimensional binary BECs. In this work, our first objective
is to produce new analytical solutions for the DWs in binary BECs,
taking into account the linear inter-conversion (Rabi coupling, RC)
between the species. The RC gives rise to effective attraction
between the immiscible components, therefore it essentially affects
the respective MIM transition and DWs generated by it, which
separate two \textit{partially mixed} components. Further, by means
of systematic numerical simulations we explore the stability of DWs
against periodic variation of the strength of the inter-component
repulsion (\textit{nonlinearity management}, NM). The results are
summarized in the form of a stability diagram for the DWs in the
parameter plane of the NM strength and frequency. Our numerical
analysis is performed, chiefly, in the ring-shaped system with
periodic boundary conditions, of a sufficiently large length, with
two DWs placed at diametrically opposite positions on the ring. The
corresponding experimental setup can be a toroidal trap with strong
radial confinement. In some cases, to explore the (in)stability of
the analytical solutions, we also use the box trap, i.e., an
infinitely deep rectangular potential well.

The subsequent presentations starts with the formulation of the
system in Section 2, which also includes the above-mentioned
estimate for the upshift of the MIM transition under the pressure of
the confining potential, as well as a newly found exact DW solution
for the system including the repulsive or attractive
P\"{o}schl-Teller potential. Section 3 reports approximate
analytical results for DWs and MIM-transition point, obtained by
means of a variational method. Comparison with numerical findings
demonstrates high accuracy of the analytical approximation. In
Section 4 we summarize findings for stable and unstable dynamical
regimes in the system including the \textit{miscibility management}
(induced by the NM). The paper is concluded by Section 5.

\section{The system with and without the trapping potential}

\subsection{The estimate of the shift of the MIM transition point under the
action of the parabolic trapping potential}

The underlying system of one-dimensional GP equations for the
component wave functions $\psi _{\pm }$ of the binary condensate is
written, in the scaled form, as
\begin{eqnarray}
i\frac{\partial \psi _{+}}{\partial t} &=&-\frac{1}{2}\frac{\partial
^{2}\psi _{+}}{\partial x^{2}}+\left( |\psi _{+}|^{2}+g|\psi
_{-}|^{2}\right) \psi _{+}+\frac{\Omega ^{2}}{2}x^{2}\psi
_{+}-\kappa \psi
_{-},  \label{with Om 1} \\
i\frac{\partial \psi _{-}}{\partial t} &=&-\frac{1}{2}\frac{\partial
^{2}\psi _{-}}{\partial x^{2}}+\left( |\psi _{-}|^{2}+g|\psi
_{+}|^{2}\right) \psi _{-}+\frac{\Omega ^{2}}{2}x^{2}\psi
_{-}-\kappa \psi _{+},  \label{with Om 2}
\end{eqnarray}%
where $\Omega ^{2}$ is the strength of the trapping quadratic
potential, and $g>0$ is the relative strength of the inter-component
repulsion, while the intra-component repulsion coefficient is scaled
to be $1$. Further, $\kappa$ is the RC constant, which is fixed
to be positive, as its negative value can be reversed by
substitution $\psi_{-}\to -\psi_{-}$. It is convenient to fix
$\kappa > 0$, as then stable two-component states, which provide for
the minimum of the RC energy, correspond to identical signs of the
components $\psi_{+,-}$.

To explore the MIM
transition threshold in the framework of this system, one can look
for a steady-state solution to Eqs. (\ref{with Om 1}) and (\ref{with
Om 2}) with chemical potential $\mu $,
\begin{equation}
\psi _{\pm }(x,t)=e^{-i\mu t}u_{\pm }(x),  \label{st}
\end{equation}%
where real functions $u_{\pm }(x)$ satisfy the system of coupled
stationary equations
\begin{eqnarray}
\frac{1}{2}\frac{d^{2}u_{+}}{dx^{2}}+u_{+}(\mu -u_{+}^{2}-gu_{-}^{2})-\frac{%
\Omega ^{2}}{2}x^{2}u_{+}+\kappa u_{-} &=&0,  \label{up} \\
\frac{1}{2}\frac{d^{2}u_{-}}{dx^{2}}+u_{-}(\mu -u_{-}^{2}-gu_{+}^{2})-\frac{%
\Omega ^{2}}{2}x^{2}u_{-}+\kappa u_{+} &=&0.  \label{um}
\end{eqnarray}%
In particular, a symmetric spatially even solution of Eqs. (\ref{up}) and (%
\ref{um}), $u_{\pm }(x)\equiv u_{\mathrm{symm}}(x)$, $u_{\mathrm{symm}%
}(-x)=u_{\mathrm{symm}}(x)$, is produced by the single equation,%
\begin{equation}
\frac{1}{2}\frac{d^{2}u_{\mathrm{symm}}}{dx^{2}}+u_{\mathrm{symm}}\left[
(\mu +\kappa )-(1+g)u_{\mathrm{symm}}^{2}\right] +\frac{\Omega ^{2}}{2}%
x^{2}u_{\mathrm{symm}}=0.  \label{symm}
\end{equation}%
Stationary states produced by Eqs. (\ref{up}) and (\ref{um}) are
characterized by the total norm,
\begin{equation}
N=\int_{-\infty }^{+\infty }\sum_{+,-}u_{\pm }^{2}(x)dx,  \label{N}
\end{equation}%
the corresponding energy of the system being
\begin{equation}
E=\int_{-\infty }^{+\infty }\left\{ \frac{1}{2}\sum_{+,-}\left[ \left( \frac{%
du_{\pm }}{dx}\right) ^{2}+u_{\pm }^{4}+\Omega ^{2}x^{2}u_{\pm
}^{2}\right] +g\left( u_{+}u_{-}\right) ^{2}-\kappa
u_{+}u_{-}\right\} dx.  \label{E}
\end{equation}

The MIM transition is initiated by a small spatially odd perturbation, $%
\delta u(-x)=-\delta u(x)$, added to $u_{\mathrm{symm}}(x)$:%
\begin{equation}
\left( u_{\pm }(x)\right)
_{\mathrm{perturbed}}=u_{\mathrm{symm}}(x)\pm \delta u(x)\text{.}
\label{delta}
\end{equation}%
Substituting this expression in expression (\ref{E}), the quartic
terms give
rise to the respective energy perturbation%
\begin{equation}
\left( \delta E\right) _{\mathrm{quart}}=2\int_{-\infty }^{+\infty }(3-2g)u_{%
\mathrm{symm}}^{2}(x)\left( \delta u(x)\right) ^{2}dx.
\label{deltaE}
\end{equation}%
Without a detailed analysis of the solution, it is easy to conclude
that other contributions to the energy perturbation, produced by the
kinetic-energy, trapping-potential, and RC terms in Eq. (\ref{E}),
are negligible in comparison to one (\ref{deltaE}) under conditions
\begin{equation}
\kappa /N\ll \sqrt{\Omega }\ll N.  \label{<<<<}
\end{equation}%
Then, also irrespective of the explicit form of functions $u_{\mathrm{symm}%
}(x)$ and $\delta u(x)$, Eq. (\ref{deltaE}) predicts the MIM
transition, i.e., the instability of the mixed state, following from
$\left( \delta E\right) _{\mathrm{quart}}<0$, at $g>3/2$. In
comparison to the above-mentioned MIM condition (\ref{aaa}), which
in the present notation takes the form of $g>1$, the expansion of
the stability region of the mixed state from $g<1$ to $g<3/2$ is a
result of the above-mentioned pressure applied to the binary BEC by
the trapping potential.

In another parameter range, \textit{viz}., $\kappa \gg \Omega $, the
effect of the trapping potential is negligible \cite{merhasin2005},
and the
boundary of the instability of the mixed state shifts from $g=1$ to $%
g=1+2\kappa /n_{0}$, where $n_{0}\equiv 2u_{\pm }^{2}(x=0)$ is the
total density of the binary BEC at the center, as shown below by Eq.
(\ref{kappa}).

\subsection{An exact DW solution supported by the P\"{o}schl-Teller potential%
}

In the special case of $g=3$ and $\Omega=0$, an exact DW solution to Eqs. (\ref{up}) and (%
\ref{um}) was found in Ref.~\cite{malomed2021}:
\begin{equation}
\left\{
\begin{array}{c}
\psi _{+}(x) \\
\psi _{-}(x)%
\end{array}%
\right\} (g=3)=\frac{1}{2}e^{-it}\left\{
\begin{array}{c}
\sqrt{1+\kappa }-\sqrt{1-\kappa }\,\mathrm{tanh}(\sqrt{1-\kappa }\,x) \\
\sqrt{1+\kappa }+\sqrt{1-\kappa }\,\mathrm{tanh}(\sqrt{1-\kappa }\,x)%
\end{array}%
\right\} ,  \label{exact1}
\end{equation}%
in which the chemical potential is fixed as $\mu =1$ [cf. Eq.
(\ref{st})] by means of scaling. This solution, which exists for
values of the RC coefficient
\begin{equation}
\kappa <\kappa _{\mathrm{crit}}\equiv 1,  \label{1}
\end{equation}%
satisfies boundary conditions
\begin{eqnarray}
u_{+}(x\rightarrow -\infty ) &=&u_{-}(x\rightarrow +\infty )=\frac{1}{2}%
\left( \sqrt{1+\kappa }+\sqrt{1-\kappa }\,\right) ,  \label{bc1} \\
u_{+}(x\rightarrow +\infty ) &=&u_{-}(x\rightarrow -\infty )=\frac{1}{2}%
\left( \sqrt{1+\kappa }-\sqrt{1-\kappa }\,\right) .  \label{bc2}
\end{eqnarray}%
Note that, for $g=3$, condition (\ref{1}) exactly agrees with the
general existence condition (\ref{kappa}), which is obtained below
as a property of approximate analytical solutions. Our numerical
analysis (not shown here in detail) \emph{corroborates the full
stability} of the DW produced by solution (\ref{exact1}) for all
values of $\kappa $ from the existence range (\ref{1}) (the
stability of this solution was not explicitly studied in Ref.
\cite{malomed2021}).

A new result reported here is the possibility to extend the exact DW
solution to the system of GP equations including the
P\"{o}schl-Teller
potential \cite{Fluegge}:%
\begin{eqnarray}
i\frac{\partial \psi _{+}}{\partial t} &=&-\frac{1}{2}\frac{\partial
^{2}\psi _{+}}{\partial x^{2}}+\left( |\psi _{+}|^{2}+g|\psi
_{-}|^{2}\right) \psi _{+}+\frac{W}{\cosh ^{2}(\alpha x)}\psi
_{+}-\kappa
\psi _{-},  \label{PT+} \\
i\frac{\partial \psi _{-}}{\partial t} &=&-\frac{1}{2}\frac{\partial
^{2}\psi _{-}}{\partial x^{2}}+\left( |\psi _{-}|^{2}+g|\psi
_{+}|^{2}\right) \psi _{-}+\frac{W}{\cosh ^{2}(\alpha x)}\psi
_{-}-\kappa \psi _{+},  \label{PT-}
\end{eqnarray}%
with specially selected values of the strength and width parameters,
$W$ and
$\alpha $:%
\begin{equation}
W=\frac{3-g}{g-1}\frac{g-1-2\kappa }{4},~\alpha =\sqrt{\frac{g-1-2\kappa }{2}%
}.  \label{Walpha}
\end{equation}%
This potential can be readily implemented in the experimental
setting for BEC with the help of an appropriately shaped laser beam
illuminating the condensate \cite{laser-beam}.

Again fixing $\mu =1$ by means of rescaling, the exact solution of Eqs. (\ref%
{PT+}) and (\ref{PT-}), with the potential's parameters chosen as per Eq. (%
\ref{Walpha}), is%
\begin{equation}
\left\{
\begin{array}{c}
\psi _{+}(x) \\
\psi _{-}(x)%
\end{array}%
\right\} =\frac{1}{2}e^{-it}\left\{
\begin{array}{c}
A-B\,\mathrm{tanh}(\alpha \,x) \\
A+B\,\mathrm{tanh}(\alpha \,x)%
\end{array}%
\right\}  \label{exact2}
\end{equation}%
(cf. Eq. (\ref{exact1})), with%
\begin{equation}
A=\sqrt{\frac{g-1+2\kappa }{g-1}},~B=\sqrt{\frac{g-1-2\kappa
}{g-1}}. \label{AB}
\end{equation}%
In the case of $g=3$, this exact solution carries over into the
above-mentioned one (\ref{exact1}).

\begin{figure}[htbp]
\centerline{ $a)$ \hspace{5cm} $b) \hspace{5cm} c)$}
\centerline{\includegraphics[width=5cm,height=4cm]{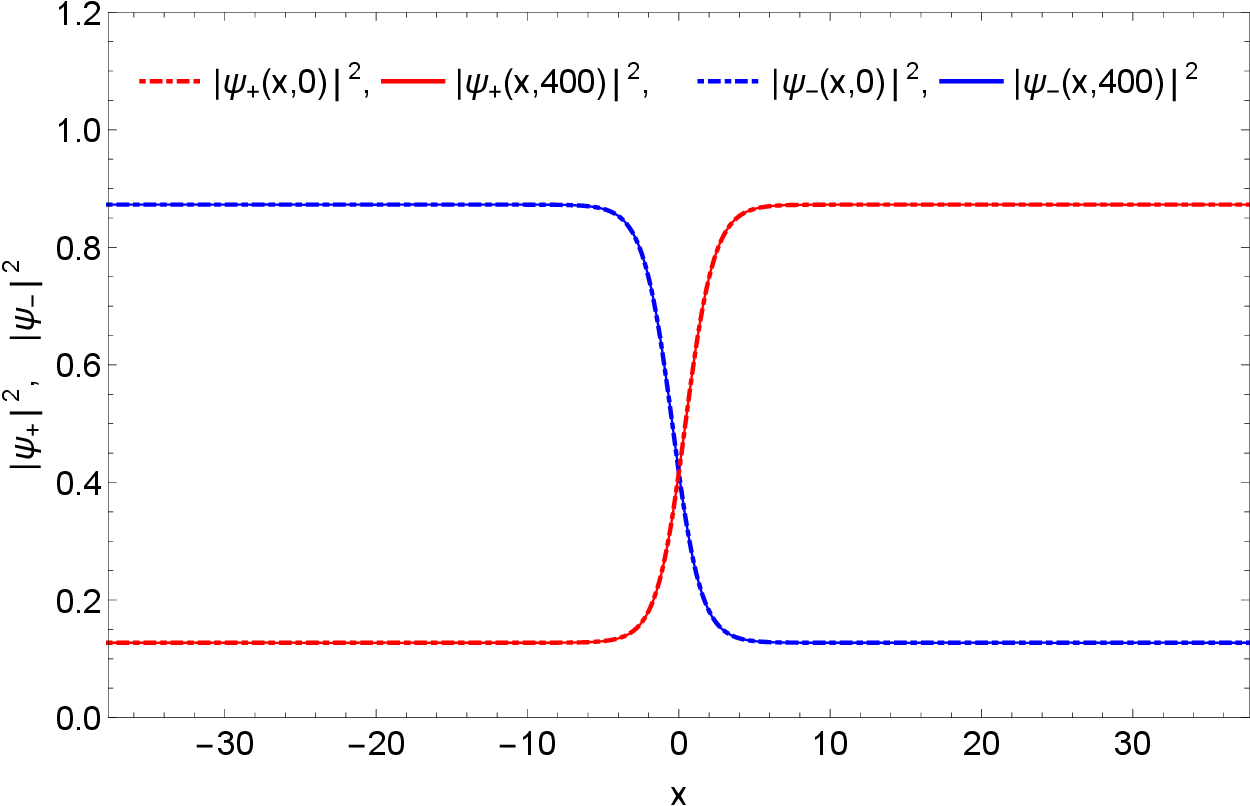}\quad
            \includegraphics[width=5cm,height=4cm]{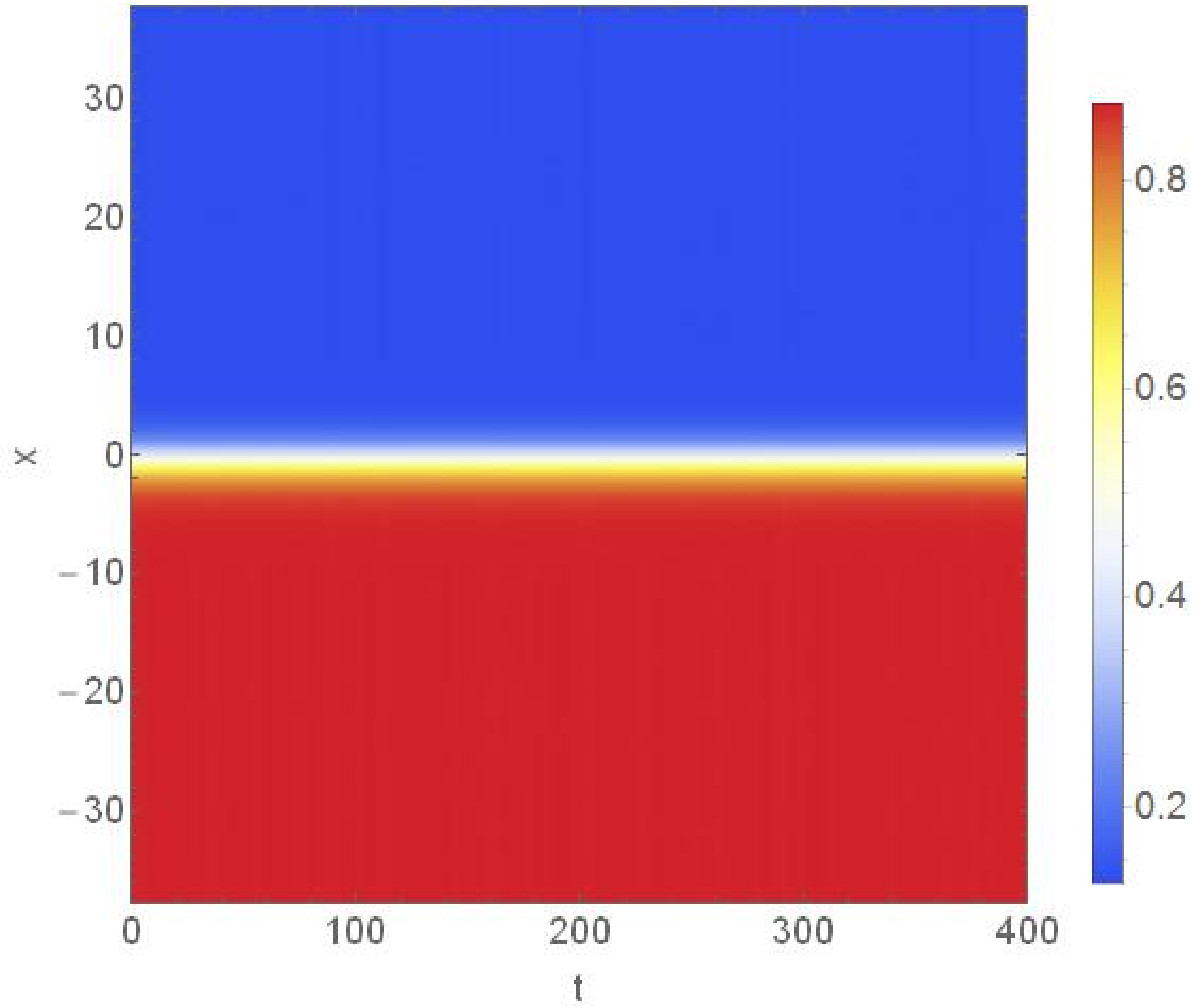}\quad
            \includegraphics[width=5cm,height=4cm]{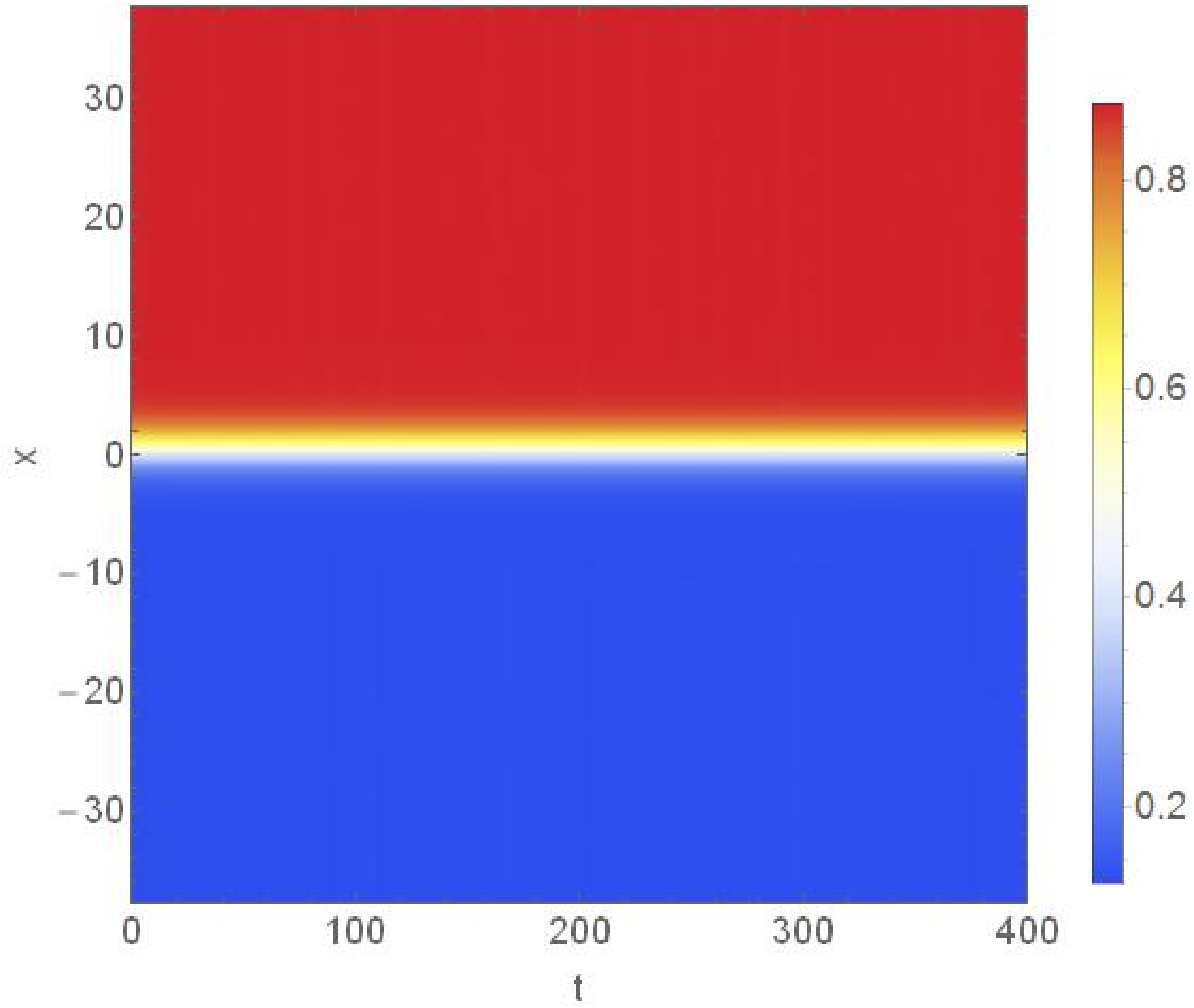}}
\centerline{ $d)$ \hspace{5cm} $e) \hspace{5cm} f)$}
\centerline{\includegraphics[width=5cm,height=4cm]{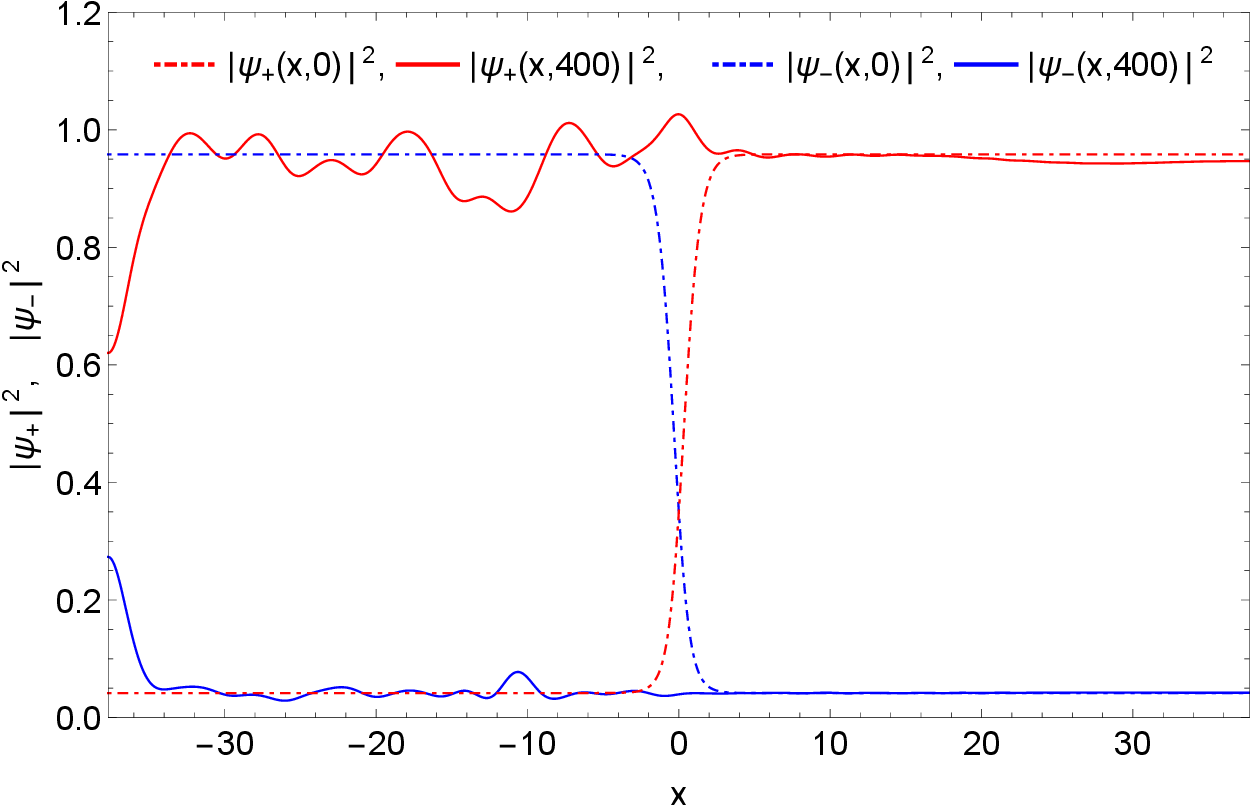}\quad
            \includegraphics[width=5cm,height=4cm]{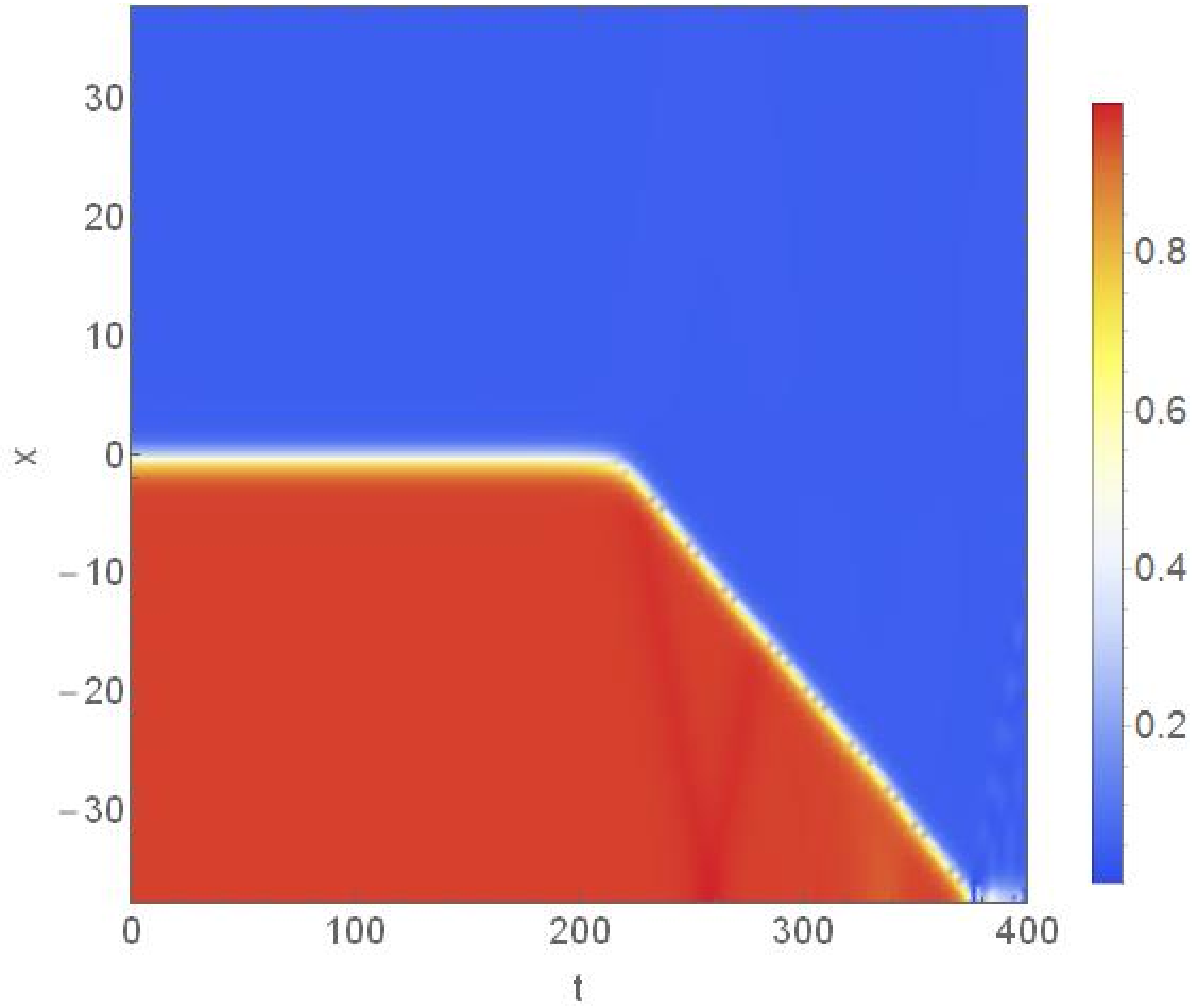}\quad
            \includegraphics[width=5cm,height=4cm]{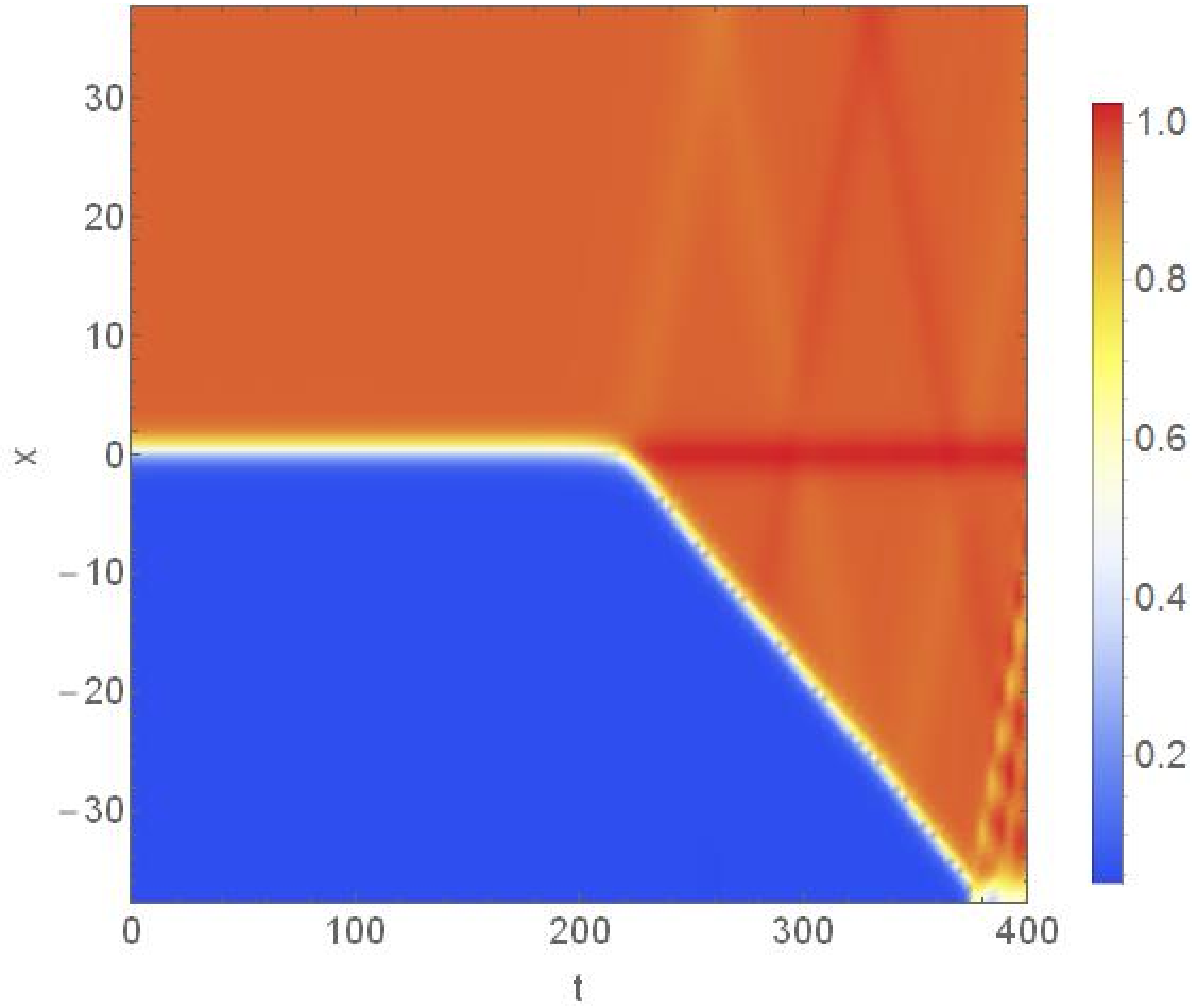}}
\caption{a), d): Snapshots of the solutions produced by the
simulations of the GP system (\protect\ref{PT+})-(\protect\ref{PT-})
with the box trap, using the exact analytical solution
(\protect\ref{exact2}) as initial conditions. The initial and final
forms of the numerical solutions are represented by the dot-dashed
and solid lines, respectively. b), c) The spatiotemporal evolution of
components $|\psi_{+}(x,t)|^2$ and $|\psi_{-}(x,t)|^2$, initiated by
the input with $\protect\kappa =0.5,\,g=2.5$ [hence $W>0$ in Eq.
(\protect\ref{Walpha}), which corresponds to the repulsive
potential], shows stable propagation in the course of long time,
$t=400$. e), f): In contrast, the solution produced by the input with
$\protect\kappa =0.5,\,g=3.5$ (hence $ W<0 $, making the potential
attractive), is unstable. The unstable DW starts to move, eventually
hitting the border of the box trap and thus disintegrating. } \label{fig1}
\end{figure}

Note that, according to Eq. (\ref{Walpha}), the solution exists in
the case of $g>1+2\kappa $, and the potential is repulsive or
attractive ($W>0$ or $W<0$, respectively), in the cases of $g<3$ or
$g>3$. In fact, the repulsive potential, which helps to separate the
two components, facilitates the creation of the DW, while the
attractive potential makes the DW solution unfavorable. In
accordance with this expectation, direct simulations of the coupled
GPE (\ref{PT+})-(\ref{PT-}) show that the repulsive potential
supports stable DW, while the attractive one leads to its
instability, as demonstrated in Fig.~\ref{fig1}. In the latter case,
the instability develops as spontaneous motion of the DW, which ends
up with destruction of the DW upon collision with the border of the
box trap, as shown by the spatiotemporal density plots in
Figs.~\ref{fig1}e),~f).

\subsection{The system without the trapping potential}

Below, we address the system which does not include the trapping
potential, but may include the action of the NM:\
\begin{eqnarray}
i\frac{\partial \psi _{+}}{\partial t} &=&-\frac{1}{2}\frac{\partial
^{2}\psi _{+}}{\partial x^{2}}+\left( |\psi _{+}|^{2}+g(t)|\psi
_{-}|^{2}\right) \psi _{+}-\kappa \psi _{-},  \label{gpe1} \\
i\frac{\partial \psi _{-}}{\partial t} &=&-\frac{1}{2}\frac{\partial
^{2}\psi _{-}}{\partial x^{2}}+\left( |\psi _{-}|^{2}+g(t)|\psi
_{+}|^{2}\right) \psi _{-}-\kappa \psi _{+},  \label{gpe2}
\end{eqnarray}%
where the NM is represented by the time-periodic modulation of the
cross-repulsion coefficient:%
\begin{equation}
g(t)=g_{0}+\varepsilon \sin (\omega t).  \label{gt}
\end{equation}%
The objective of the NM is to introduce \textquotedblleft
miscibility management" at frequency $\omega $. In particular, this
case is relevant if the constant part $g_{0}$ of the cross-repulsion
coefficient $g_{0}$ is slightly greater than the critical value
$g_{\mathrm{MIM}}$ at the
MIM-transition point, and the NM\ amplitude is small, $\varepsilon \ll g_{%
\mathrm{MIM}}$.

The Lagrangian of system (\ref{gpe1})-(\ref{gpe2}) is
\begin{equation}
L=\int_{-\infty }^{+\infty }\mathcal{L}dx,
\end{equation}%
with the Lagrangian density
\begin{eqnarray}
\mathcal{L} &=&\sum_{+,-}\left[ \left( \frac{i}{2}\psi _{\pm }^{\ast }\frac{%
\partial \psi _{\pm }}{\partial t}+\mathrm{c.c.}\right) -\frac{1}{2}%
\left\vert \frac{\partial \psi _{\pm }}{\partial x}\right\vert ^{2}-\frac{1}{%
2}\left\vert \psi _{\pm }\right\vert ^{4}\right]  \notag \\
&&-g(t)\,|\psi _{+}|^{2}|\psi _{-}|^{2}+\kappa (\psi _{+}^{\ast }\psi _{-}+%
\mathrm{c.c.}),  \label{Ld}
\end{eqnarray}%
where both $\ast $ and $\mathrm{c.c.}$ stand for the complex
conjugate. Below, the Lagrangian is employed to derive a variational
approximation for dynamical DW states.

\section{The sine-Gordon (SG) approximation}

\subsection{The general analysis}

In the case of the proximity to the MIM transition, a natural ansatz
for an approximate solution is adopted as
\begin{equation}
\left\{
\begin{array}{c}
\psi _{+} \\
\psi _{-}%
\end{array}%
\right\} =\sqrt{n_{0}}\exp (-in_{0}t)\left\{
\begin{array}{c}
(\mathrm{cos}\,\chi )\exp (+i\theta ) \\
(\mathrm{sin}\,\chi )\exp (+i\theta )%
\end{array}%
\right\} ,  \label{ansatz}
\end{equation}%
where $n_{0}$ is the background amplitude (eventually, one may set $%
n_{0}\equiv 1$ by means of rescaling), while $\chi (x,t)$ and
$\theta (x,t)$ are real phases, equations for which should be
derived with the help of the Lagrangian representation of the
system. For the stationary configurations, ansatz (\ref{ansatz})
coincides with the one introduced long ago in Ref.
\cite{malomed1994}, which has $\theta =0$.

The substitution of ansatz (\ref{ansatz}) in the Lagrangian density
(\ref{Ld}) yields the corresponding effective Lagrangian density:
\begin{eqnarray}
&&\mathcal{L}_{\mathrm{effect}}=-\frac{n_{0}}{2}\,\mathrm{cos}(2\chi )\,%
\frac{\partial \theta }{\partial t}-\frac{n_{0}}{2}\left[ \left( \frac{%
\partial \chi }{\partial x}\right) ^{2}+\left( \frac{\partial \theta }{%
\partial x}\right) ^{2}\right]  \notag \\
&&-\frac{n_{0}^{2}}{8}[(3+g(t))+(1-g(t))\,\mathrm{cos}(4\chi
)]+n_{0}\kappa \sin (2\chi )\,\mathrm{cos}(2\theta ).  \label{Leff}
\end{eqnarray}%
The Euler-Lagrange equations for $\chi $ and $\theta$ are produced
by the application of the variational procedure to the Lagrangian
produced by the effective density (\ref{Leff}):
\begin{gather}
\mathrm{sin}(2\chi )\frac{\partial \theta }{\partial
t}+\frac{\partial ^{2}\chi }{\partial
x^{2}}+\frac{n_{0}}{2}\,(1-g(t))\,\mathrm{sin}(4\chi )+2\,\kappa
\,\mathrm{cos}(2\chi )\,\mathrm{cos}(2\,\theta )=0, \label{chi1}
\\
-\mathrm{sin}(2\chi )\frac{\partial \chi }{\partial
t}+\frac{\partial
^{2}\theta }{\partial x^{2}}-2\,\kappa \,\mathrm{sin}(2\chi )\,\mathrm{sin}%
(2\,\theta )=0.  \label{theta1}
\end{gather}

In the case of the static configuration, with $g=\mathrm{const}$, it
is possible to drop $\theta $ from Eqs. (\ref{chi1}) and
(\ref{theta1}), the remaining stationary version of Eq. (\ref{chi1})
being, essentially, a stationary double-SG equation
\cite{Campbell,CSF}:
\begin{equation}
\frac{\partial ^{2}\chi }{\partial x^{2}}+\frac{n_{0}}{2}\,(1-g)\,\mathrm{sin%
}(4\chi )+2\,\kappa \,\mathrm{cos}(2\chi )=0.  \label{dsg}
\end{equation}%
In particular, if $\kappa =0$, the obvious kink solution of the SG equation (%
\ref{dsg}) corresponds to the DW, in terms of the underlying system (\ref%
{gpe1})-(\ref{gpe2}), provided that the system is immiscible, i.e.,
$g-1>0$:
\begin{equation}
\chi _{\mathrm{DW}}=\mathrm{arctan}\left( \exp \left( \sigma \sqrt{%
2\,n_{0}(g-1)}\,x\right) \right) ,  \label{chidw}
\end{equation}%
where $\sigma =\pm 1$ is the kink's polarity.

\begin{figure}[htbp]
\centerline{\includegraphics[width=8cm,height=6cm]{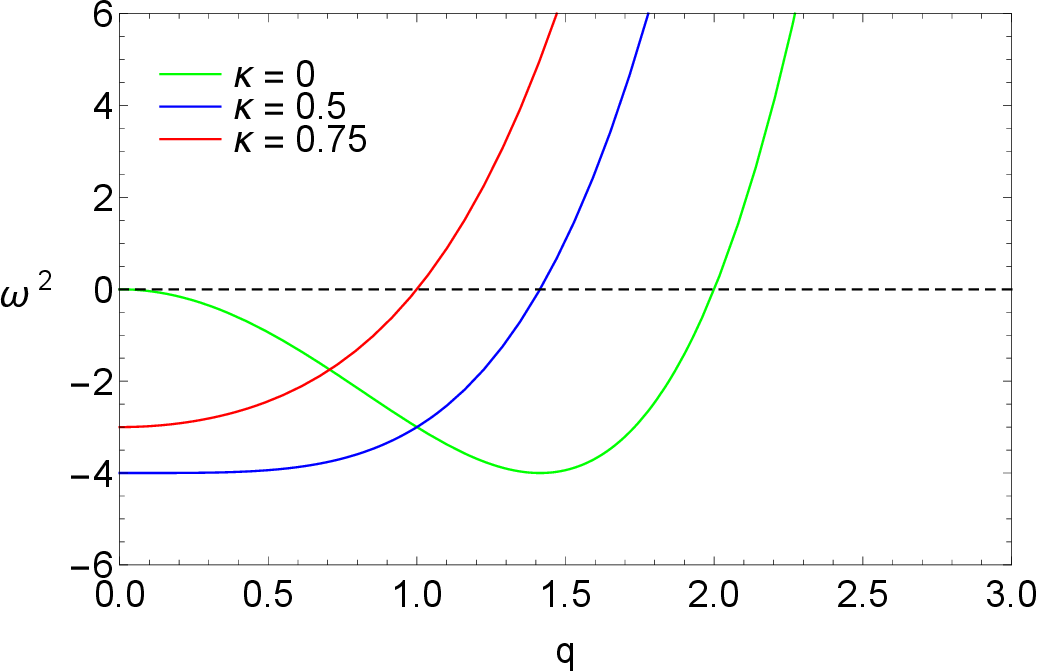}}
\caption{The dispersion relation (\protect\ref{omega2}) for the
modulational perturbations in the case of $n_{0}=1,g=3$ and
different values of the RC
coefficient $\protect\kappa $. The instability area ($\protect\omega ^{2}<0$%
) is shifted towards smaller wave numbers ($q$) as $\protect\kappa $
increases. } \label{fig2}
\end{figure}

In the case of $g=\mathrm{const}$, Eqs. (\ref{chi1}) and
(\ref{theta1}) can be used to check the modulational stability of
the uniformly mixed state, which corresponds to $\chi =\pi /4$ and
$\theta =0$, in the case of $g<1$ (miscibility), and the
modulational instability of the same state in the case of $g>1$
(immiscibility). Indeed, taking the perturbed solution as
\begin{equation}
\chi =\pi /4+\delta \chi \left( x,t\right) \quad \mathrm{and}\quad
\theta \left( x,t\right) ,  \label{chi0}
\end{equation}%
where $\delta \chi $ and $\theta $ are small perturbations, the
linearization of Eqs. (\ref{chi1}) and (\ref{theta1}) for the
perturbations yields the system
\begin{gather}
\theta _{t}+(\delta \chi )_{xx}+2\,n_{0}\,(1-g)\,\delta \chi
-4\kappa \cdot
\delta \chi =0,  \notag \\
-(\delta \chi )_{t}+\theta _{xx}-4\,\kappa \,\theta =0.
\label{pert}
\end{gather}%
The substitution of
\begin{equation}
\delta \chi \left( x,t\right) ,~\theta \left( x,t\right) \sim \exp
(iqx-i\omega t)  \label{dchi}
\end{equation}%
in Eqs. (\ref{pert}) gives rise to the dispersion relation for the
modulational perturbations:
\begin{equation}
\omega ^{2}=2n_{0}\,(1-g)(q^{2}+4\kappa )+(q^{2}+4\kappa )^{2}.
\label{omega2}
\end{equation}%
Obviously, Eq. (\ref{omega2}) gives rise to the stability and
instability in the cases of $1-g>0$ and $1-g<0$, respectively, as
expected. This dispersion relation is plotted in Fig. \ref{fig2} for
different values of the RC parameter $\kappa $.

\subsection{The effect of the RC (Rabi coupling) and comparison of the SG
approximation with numerical results}

In the case of $g(t)=\mathrm{const}$ and $\theta =0$, the double-SG
equations (\ref{dsg}) can be written as follows:
\begin{eqnarray}
\frac{d^{2}\tilde{\chi}}{dz^{2}} &=&\mathrm{sin}\tilde{\chi}-K\,\mathrm{cos}%
\left( \frac{\tilde{\chi}}{2}\right) \equiv -\frac{dU_{\mathrm{eff}}}{d%
\tilde{\chi}},  \label{dsg2} \\
U_{\mathrm{eff}}
&=&\mathrm{cos}\,\tilde{\chi}+2\,K\,\mathrm{sin}\left(
\frac{\tilde{\chi}}{2}\right) ,  \label{efpot}
\end{eqnarray}%
where $\tilde{\chi}\equiv 4\chi $, maxima of the effective potential (\ref%
{efpot}) are attained at $\tilde{\chi}_{0}=\pi +4\pi n$, and
\begin{equation}
K\equiv 4\kappa /(n_{0}\,(g-1)),  \label{K}
\end{equation}%
\begin{equation}
z\equiv \sqrt{2n_{0}(g-1)}\,x.  \label{z}
\end{equation}%
Fixed-point solutions of Eq. (\ref{dsg2}), $\tilde{\chi}=\tilde{\chi}_{0}=%
\mathrm{const}$, are ones with
\begin{equation}
\sin \left( \frac{\tilde{\chi _{0}}}{2}\right) =\frac{K}{2},
\label{fp}
\end{equation}%
and $\mathrm{cos}(\tilde{\chi}/2)=0$, i.e., $\tilde{\chi}_{0}=\pi
+2\pi n$,
where $n$ is an arbitrary integer. The fixed point (\ref{fp}) exists at $%
K\leq 2$, i.e., if the RC strength takes values below the critical
value:
\begin{equation}
\kappa <\kappa _{\mathrm{crit}}=\frac{1}{2}n_{0}\,(g-1).
\label{kappa}
\end{equation}

\begin{figure}[htbp]
\centerline{ $a)$ \hspace{5cm} $b) \hspace{5cm} c)$}
\centerline{\includegraphics[width=5cm,height=5cm]{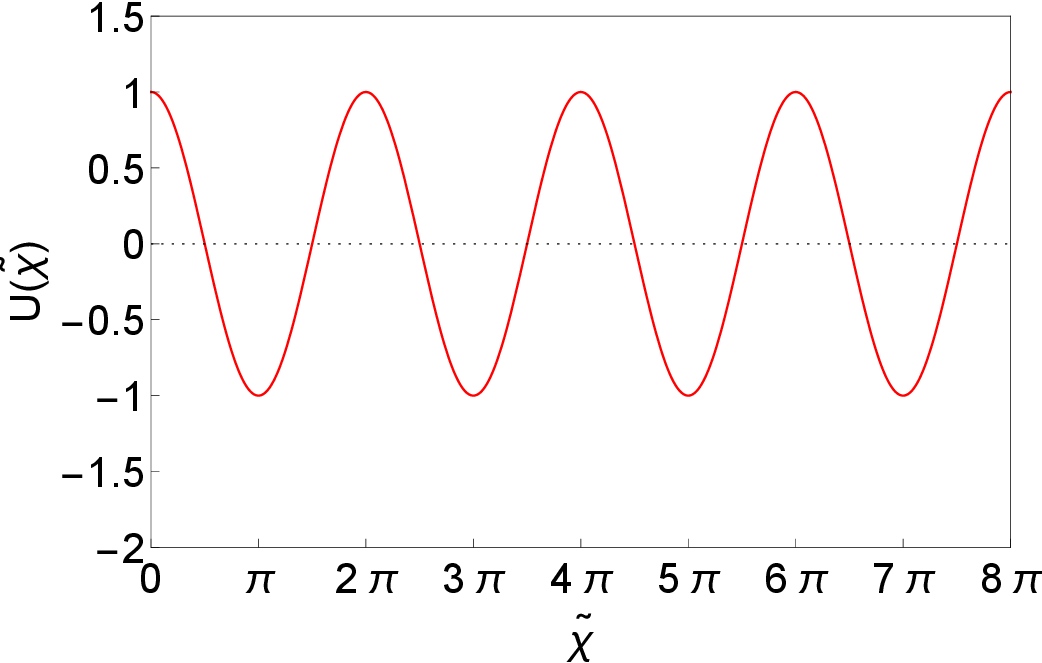}\quad
            \includegraphics[width=5cm,height=5cm]{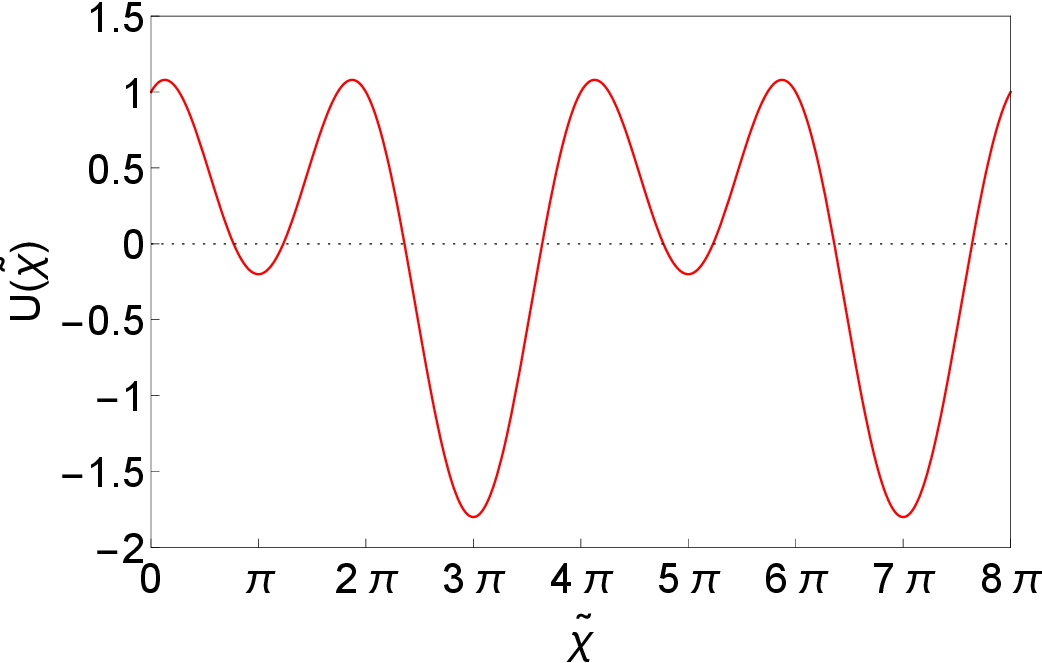}\quad
            \includegraphics[width=5cm,height=5cm]{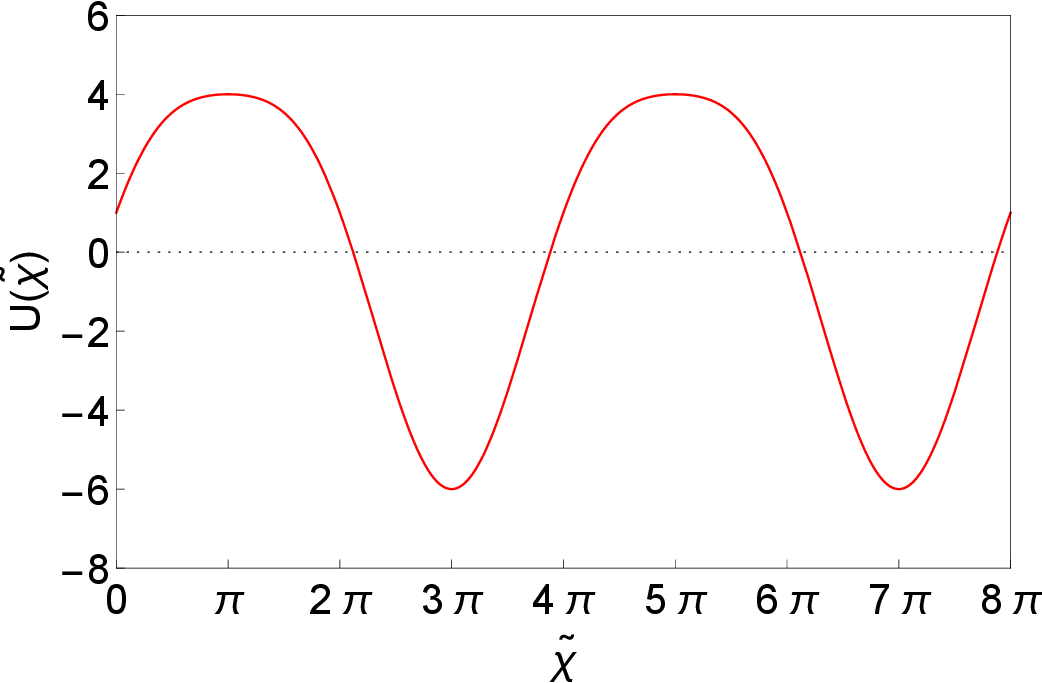}}
\caption{The effective potential Eq. (\protect\ref{efpot}) for
parameter values $n_{0}=1$, $g=3$ and a) $K=0$, b) $K=0.4$, c)
$K=2.5$.} \label{fig3}
\end{figure}

For the illustration of kink solutions of Eq. (\ref{dsg2}), it is
instructive to look at plots of the effective potential
(\ref{efpot}) produced for $K<2$ (Figs. \ref{fig3} a,b) and $K>2$
(Fig. \ref{fig3} c)).
If condition (\ref{kappa}) holds, Fig. \ref{fig3} b) suggests the
existence of two species of DWs, \textquotedblleft narrow" and
\textquotedblleft broad" ones. To obtain the respective solutions in
the analytical form, we note that, for DW solutions, Eqs.
(\ref{dsg2}) and (\ref{efpot}) can be integrated once and thus
reduced to the first-order one:
\begin{equation}
\frac{d\tilde{\chi}}{dz}=\pm \left( K-2\sin \left( \frac{\tilde{\chi}}{2}%
\right) \right) .  \label{chi2}
\end{equation}%
Then, elementary exact solutions of Eq. (\ref{chi2}) for the narrow
and broad DWs are
\begin{eqnarray}
\tilde{\chi}_{\mathrm{DW}}^{\mathrm{(narrow)}} &=&\pi \mp 4\,\mathrm{arctan}%
\left( \sqrt{\frac{1-\frac{K}{2}}{1+\frac{K}{2}}}\,\mathrm{tanh}\left( \sqrt{%
1-\frac{K^{2}}{4}}\frac{z}{2}\right) \right) ,  \label{narrow} \\
\tilde{\chi}_{\mathrm{DW}}^{\mathrm{(broad)}} &=&3\pi \pm 4\,\mathrm{arctan}%
\left( \sqrt{\frac{1+\frac{K}{2}}{1-\frac{K}{2}}}\,\mathrm{tanh}\left( \sqrt{%
1-\frac{K^{2}}{4}}\frac{z}{2}\right) \right) .  \label{broad}
\end{eqnarray}%
The solutions given by Eqs. (\ref{narrow}) and (\ref{chidw})
coincide in the absence of the RC term, $\kappa =0$. If condition
(\ref{kappa}) does not hold, Fig.~\ref{fig3} c) suggests that Eq.
(\ref{dsg}) admits a kink solution, which connects states with $\chi
=\pi /4$ and $\chi =5\pi /4$. However, as it follows from Eq.
(\ref{ansatz}), this kink is \emph{not a DW}.

\begin{figure}[htbp]
\centerline{ $a)$ \hspace{5cm} $b) \hspace{5cm} c)$}
\centerline{\includegraphics[width=5cm,height=4cm]{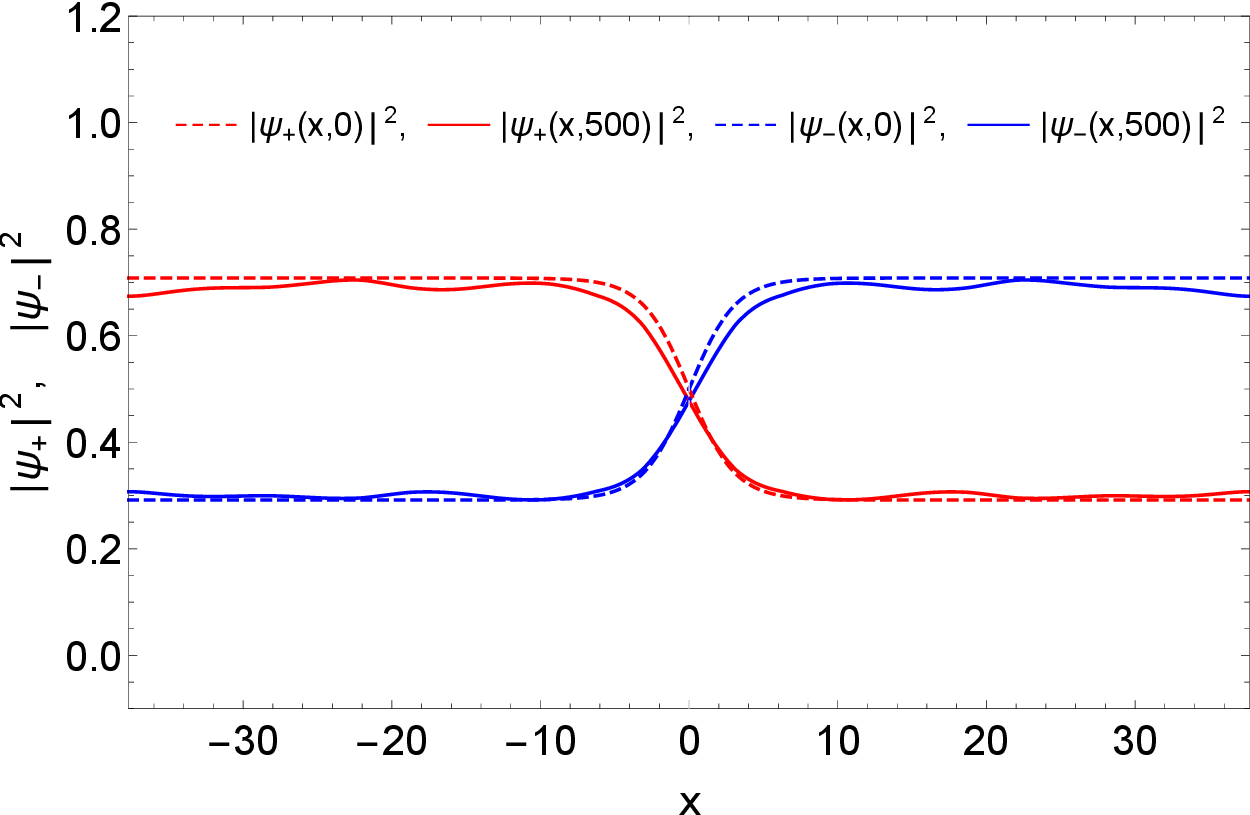}\quad
            \includegraphics[width=5cm,height=4cm]{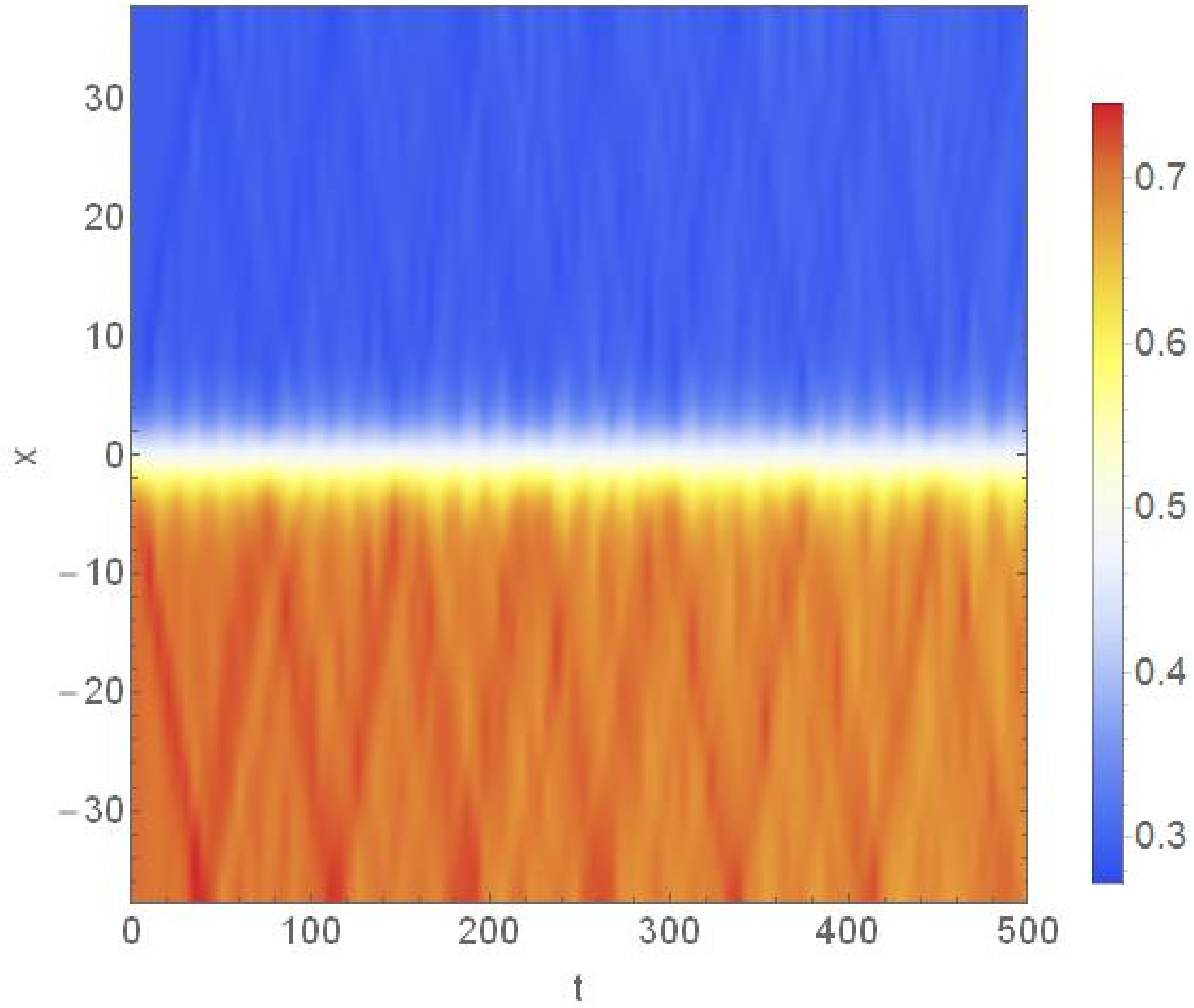}\quad
            \includegraphics[width=5cm,height=4cm]{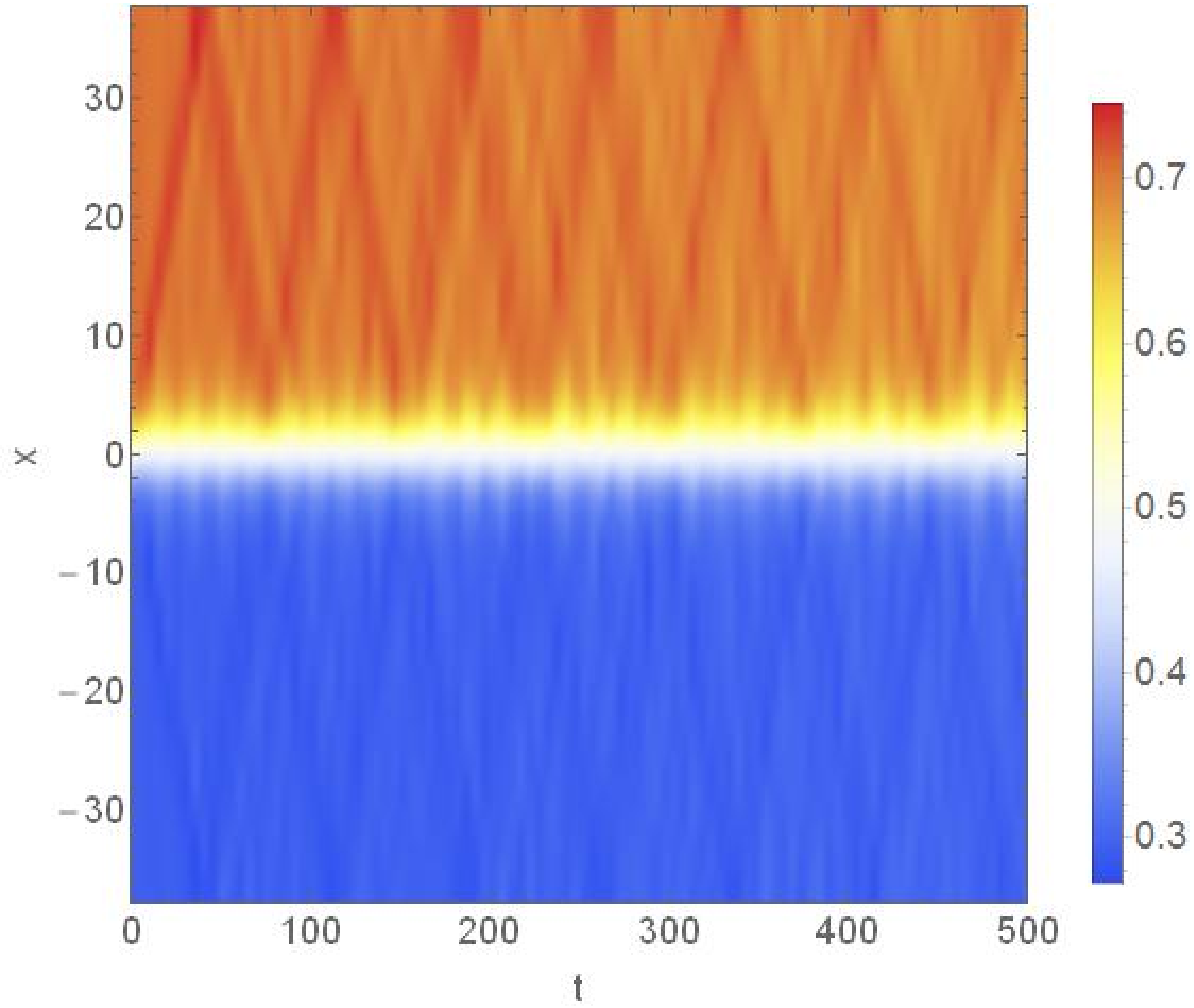}}
\centerline{ $d)$ \hspace{5cm} $e) \hspace{5cm} f)$}
\centerline{\includegraphics[width=5cm,height=4cm]{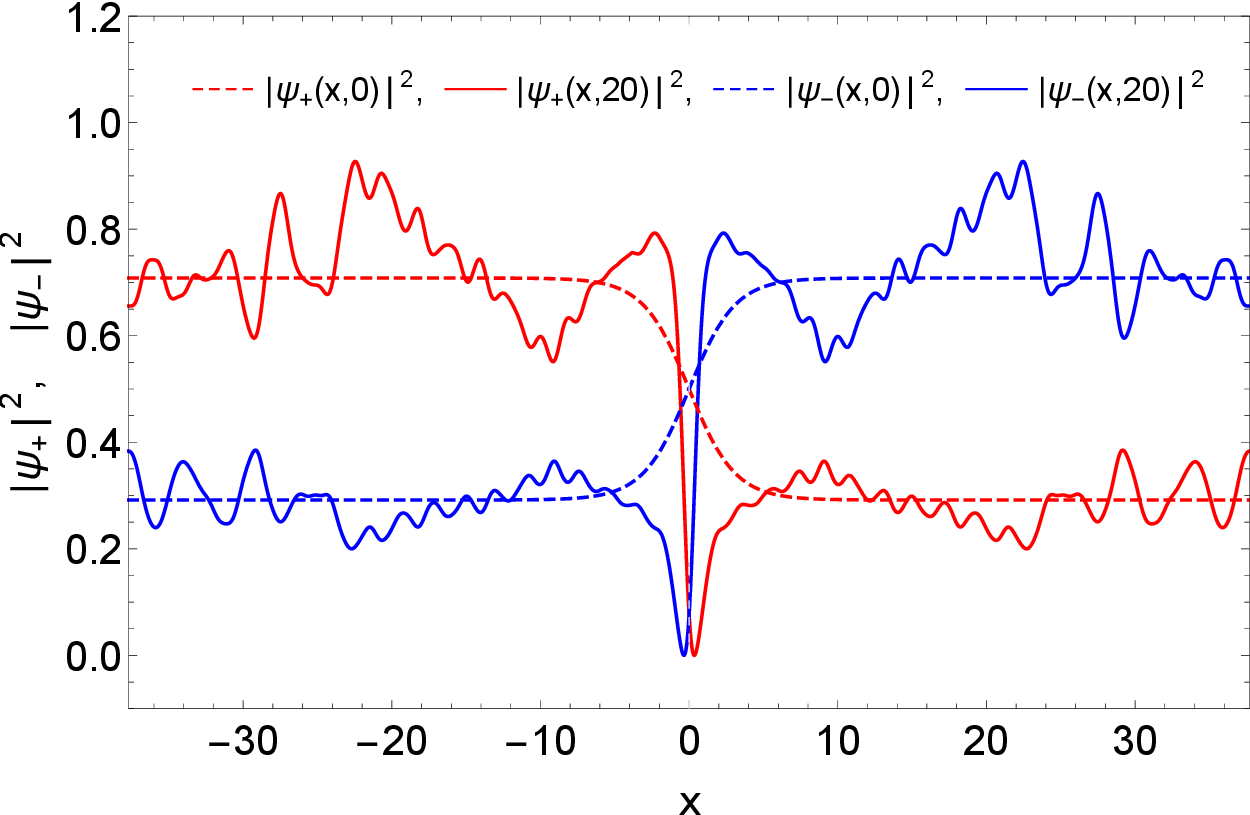}\quad
            \includegraphics[width=5cm,height=4cm]{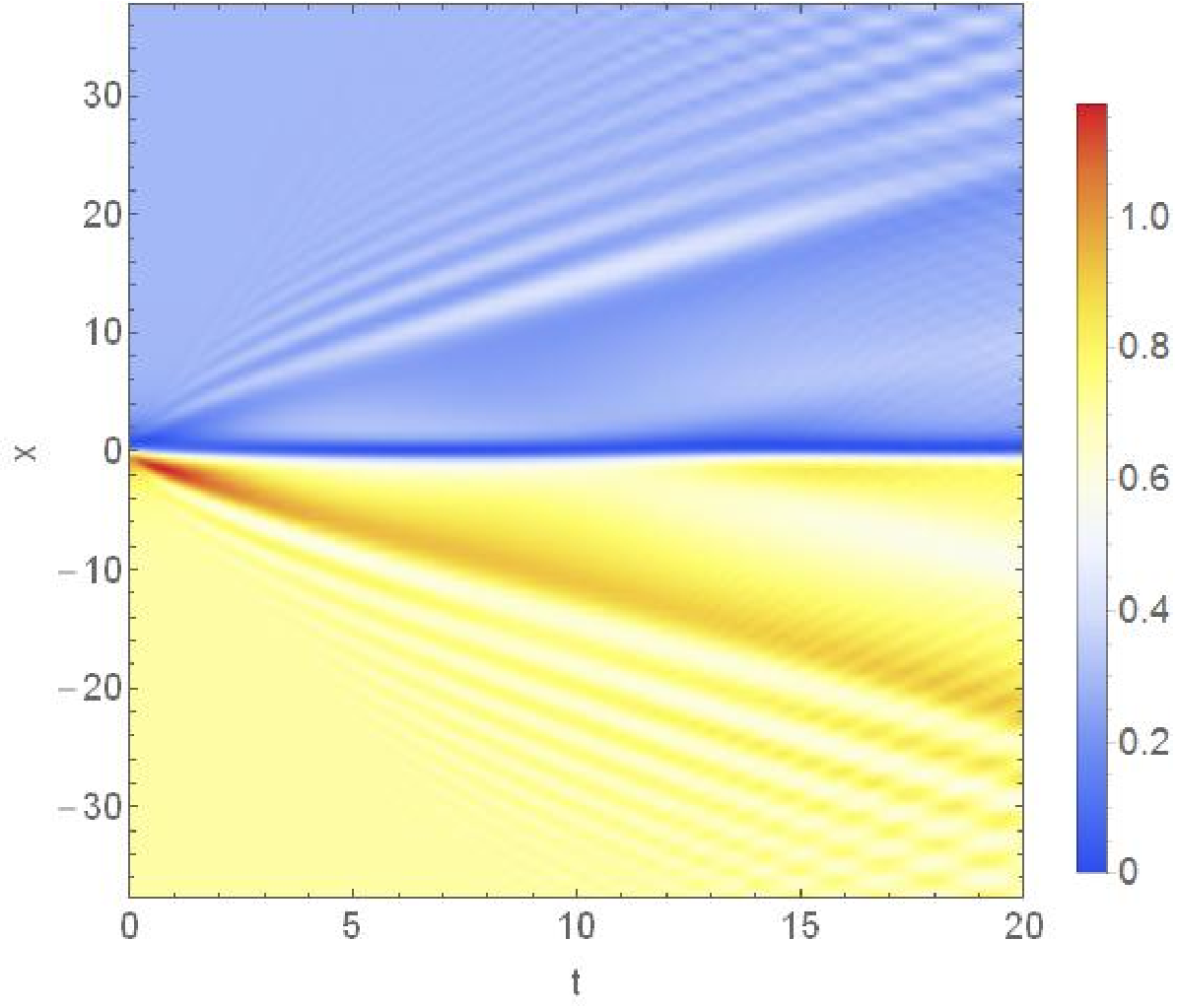}\quad
            \includegraphics[width=5cm,height=4cm]{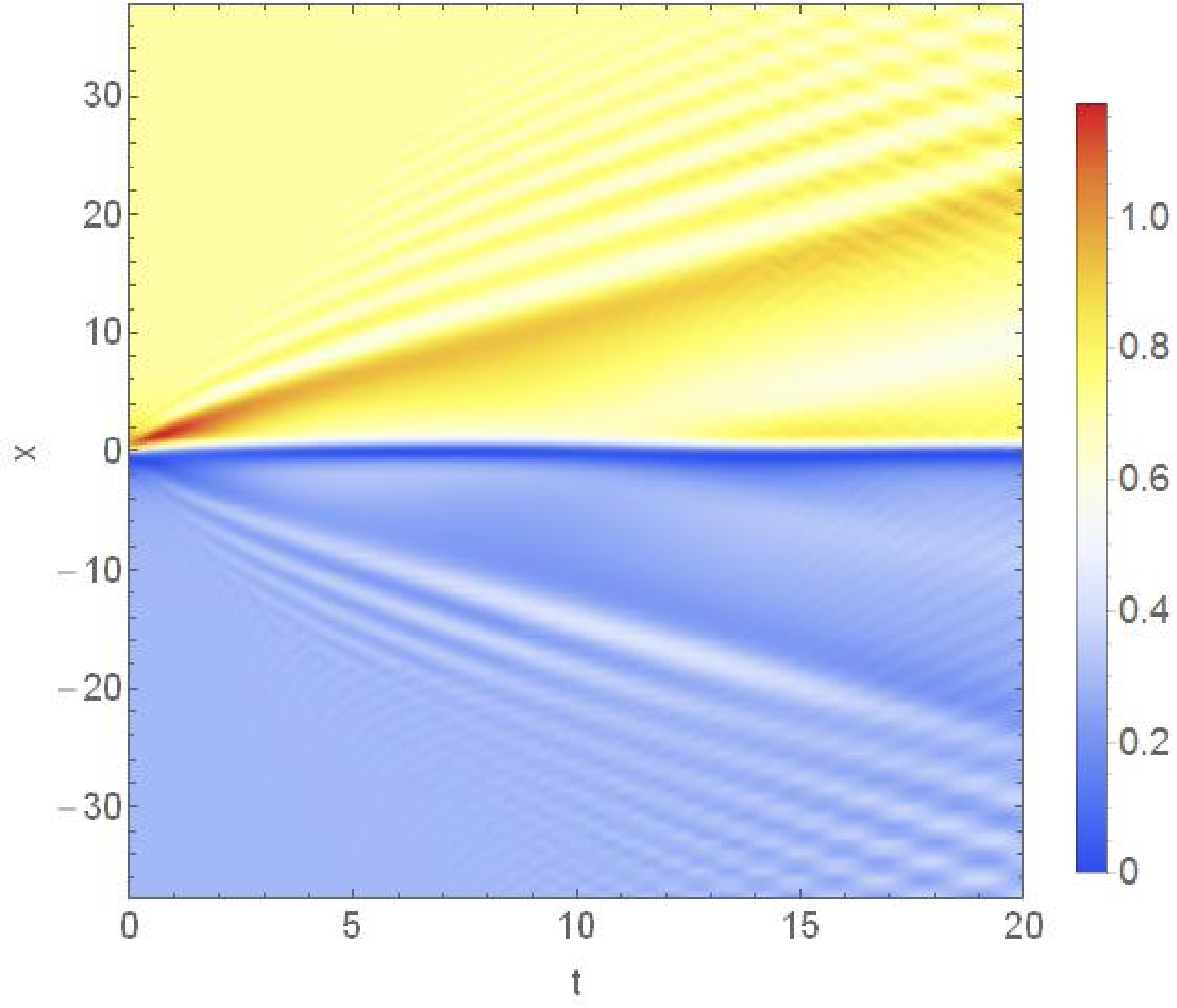}}
\caption{a), d): Snapshots of the solutions produced by the
simulations of the GP system
(\protect\ref{gpe1})-(\protect\ref{gpe2}) for the box trap with the
approximate analytical solutions (\protect\ref{narrow}) and
(\protect\ref{broad}) used as initial conditions. The input and
output of the numerical solutions are represented by the dashed and
solid lines, respectively. a) The solution initiated by
input~(\protect\ref{narrow}) shows an effectively stable evolution over
a long time, $t=500$. d) In contrast, the solution produced by input
(\protect\ref{broad}) is unstable, breaking down very quickly, by
$t=20$. The density plots b), c) and e), f) show the corresponding
spatiotemporal evolution of the components $|\psi_{+}(x,t)|^2$ and
$|\psi_{-}(x,t)|^2$. In this case, the parameters are
$n_{0}~=~1$,$\,\ g~=~2.1$,$\,\ \protect\kappa ~=~0.5$,\thinspace $\
$and $K~=~1.82$.} \label{fig4}
\end{figure}

The accuracy and stability of the above approximate solutions can be
verified by plugging them as initial conditions for the underlying
GPE system. Figure \ref{fig4} illustrates the time evolution of the\
inputs
corresponding to the approximate analytical solutions (\ref{narrow}) and (%
\ref{broad}), as produced by the simulations of Eqs.
(\ref{gpe1})-(\ref{gpe2}), using the following initial and boundary
conditions (corresponding to the
box trap, which confines the integration domain to $x\in \lbrack -L/2,+L/2]$%
, with the gradients of the wave functions at the edges of the box
fixed to be the same as given by the approximate solution):
\begin{equation}
\psi _{\pm }(x,0)=\left\{
\begin{array}{c}
\mathrm{cos}\,\chi _{\mathrm{DW}}(x) \\
\mathrm{sin}\,\chi _{\mathrm{DW}}(x)%
\end{array}%
\right\} ,~\left. \frac{\partial \psi _{\pm }(x,t)}{\partial
x}\right\vert _{x\rightarrow \pm L/2}=\frac{d}{dx}\left. \left\{
\begin{array}{c}
\mathrm{cos}\,\chi _{\mathrm{DW}}(x) \\
\mathrm{sin}\,\chi _{\mathrm{DW}}(x)%
\end{array}%
\right\} \right\vert _{x\rightarrow \pm L/2}.  \label{ibc}
\end{equation}%

\begin{figure}[htbp]
\centerline{\qquad $a)$ \hspace{6cm} $b)$}
\centerline{\includegraphics[width=6cm,height=5cm]{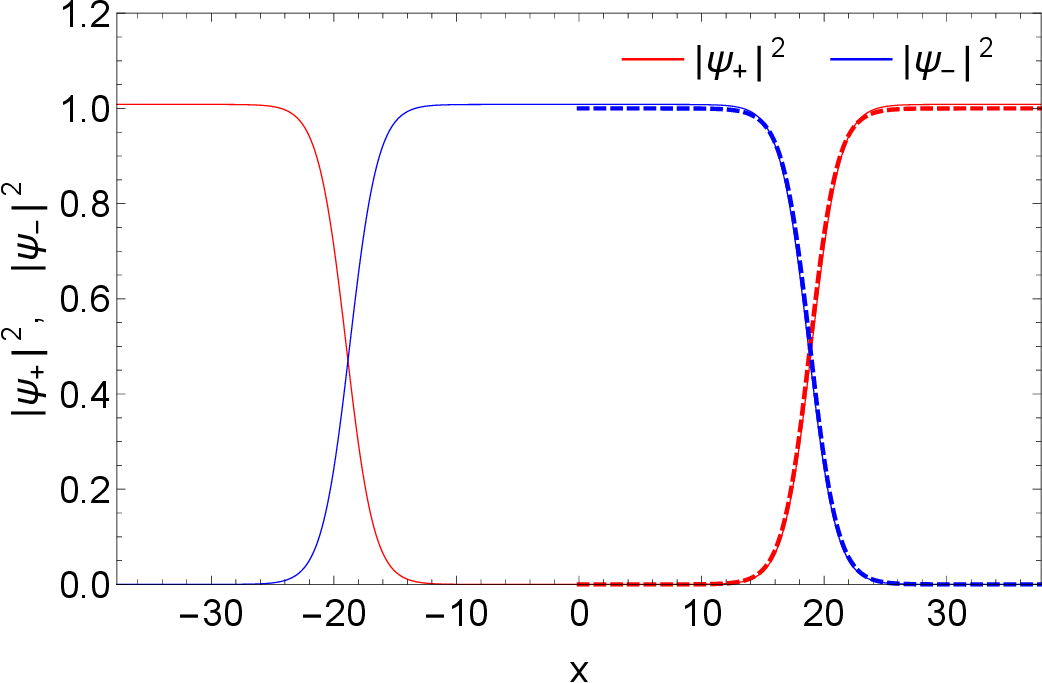}\qquad
            \includegraphics[width=6cm,height=5cm]{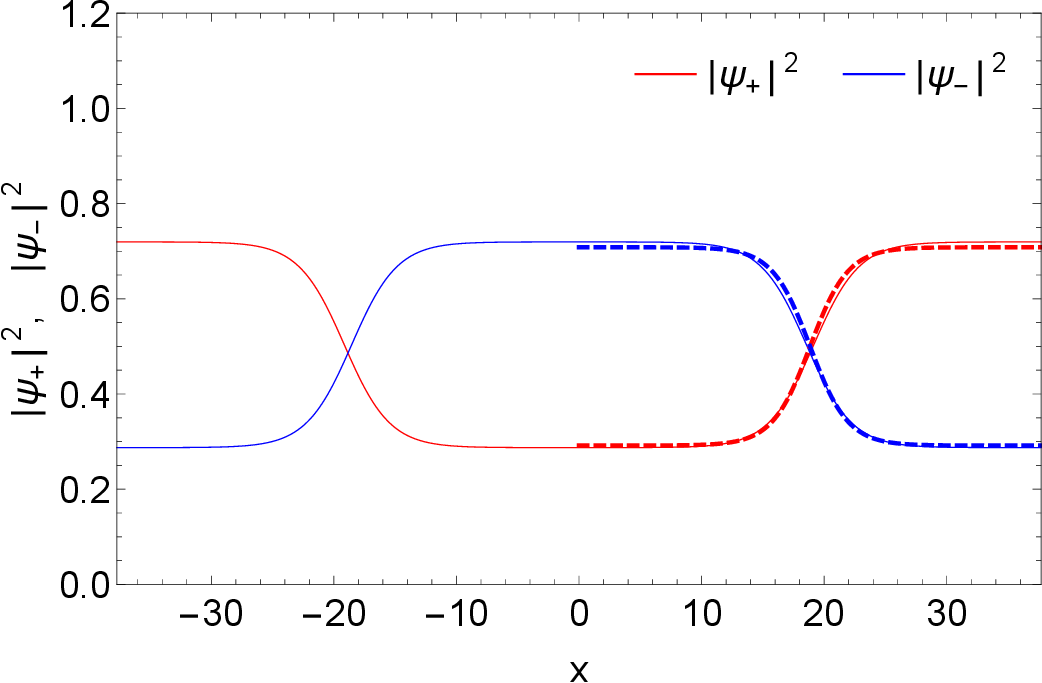}}
\caption{Component densities $|\protect\psi _{+}(x)|^{2}=\left\vert \mathrm{%
sin}\,\left( \protect\chi _{\mathrm{DW}}(x)\right) \right\vert ^{2}$ and $|%
\protect\psi _{-}(x)|^{2}=\left\vert \mathrm{cos}\,\left( \protect\chi _{%
\mathrm{DW}}(x)\right) \right\vert ^{2}$ according to the
approximate analytical solution (\protect\ref{narrow}) (dashed
lines) and its numerically obtained counterpart in the ring
configuration (solid lines) for
the following parameter sets: a) $n_{0}=1,\,g=1.1,\,\protect\kappa =0$; b) $%
n_{0}=1,\,g=2.1,\,\protect\kappa =0.5$. With these parameter values,
the
system is close to the MIM transition, as specified by Eq. (\protect\ref%
{kappa}). The approxinate solution slightly underestimates the width
of the DW interface relative to the numerical solution.}
\label{fig5}
\end{figure}

For the direct comparison of the approximate solutions with their
numerically found counterparts, we have produced the ground-state
solution of the system (\ref{gpe1})-(\ref{gpe2}) on the ring, in the
immiscibility regime, using the imaginary-time evolution method
\cite{chiofalo2000}. As said above, the ring geometry gives rise to
the state with a pair of DWs located at diametrically opposite
positions. In Fig. \ref{fig5}, the component densities $|\psi _{\pm
}(x)|^{2}$ corresponding to the approximate analytical and numerical
solutions are juxtaposed. The figure clearly shows high accuracy of
the analytical approximation.

\section{Numerical simulations of the miscibility management}

To explore the response of the DW structures to the application of
the NM,
defined as per Eq. (\ref{gt}), we numerically solved the system of Eqs. (\ref%
{gpe1}), (\ref{gpe2}), using either the approximate analytical
solutions given by Eqs. (\ref{narrow}) and (\ref{exact1}), or
numerically constructed solutions for the constant cross-repulsion
coefficient, $g=g_{0}$, as initial conditions. For the ring-shaped
system, the simulations were performed in real time by means of the
split-step fast Fourier transform \cite{adhikari2002}, with the
periodic boundary conditions,
$\psi_{\pm}(x=-L/2,t)=\psi_{\pm}(x=+L/2,t)$. For the box trap, where
the periodic boundary conditions do not apply, the standard
\texttt{NDSolve} routine of the Mathematica package has been
employed, using the initial and boundary conditions as given by Eq.
(\ref{ibc}).

Below we consider the application of the NM in two cases: far from
the MIM transition point ($g=3$) and close to it ($g=2.1$, as
$g_{\mathrm{cr}}=2$ in this case) in the ring geometry. The
perimeter of the ring is $L=24\pi $, with centers of the two DWs
initially placed at
\begin{equation}
x_{\mathrm{center}}=\pm L/4.  \label{L}
\end{equation}%
In the former case ($g=3)$, our objective is to reveal the nonlinear
resonance in the dynamics of the number of atoms on one side of the
DW, driven by weak NM. In the latter case ($g=2.1)$, we are
interested in observing the transition of the binary condensate from
an immiscible to miscible state under the action of stronger NM,
across the critical strength of the cross-repulsion coefficient,
$g$. We fix a parameter set $\left( n_{0}=1,\,\kappa =0.5\right) $,
for which $g_{\mathrm{cr}}=2$, according to Eq. (\ref{kappa}).

\subsection{The weak NM\ regime}

\begin{figure}[htbp]
\centerline{ $a)$ \hspace{5cm} $b) \hspace{6cm} c)$}
\centerline{\includegraphics[width=6cm,height=5cm]{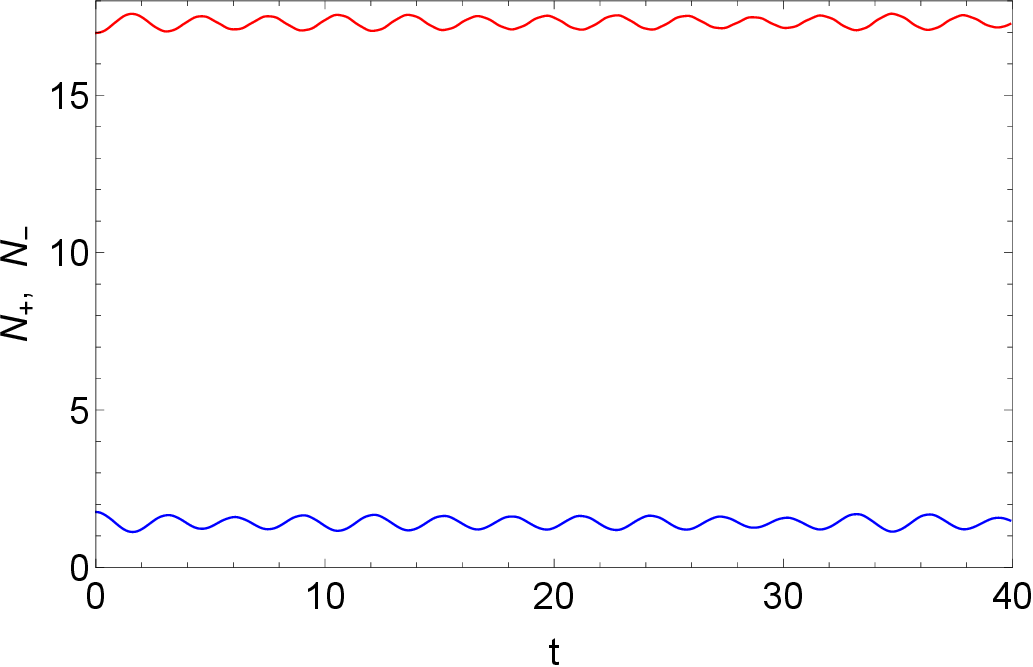}
            \includegraphics[width=6cm,height=5cm]{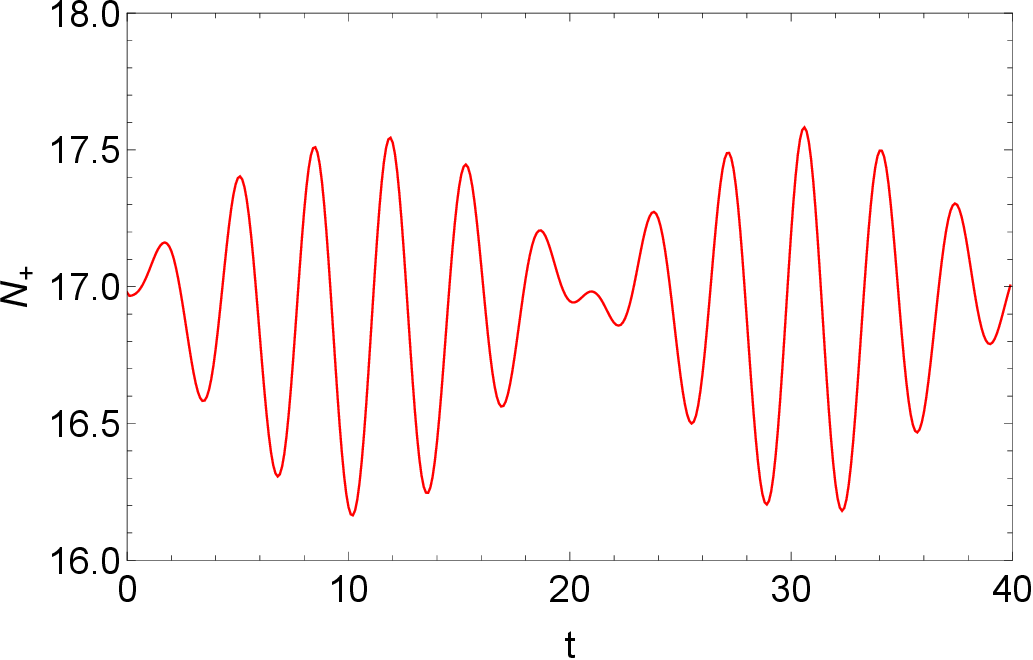}
            \includegraphics[width=6cm,height=5cm]{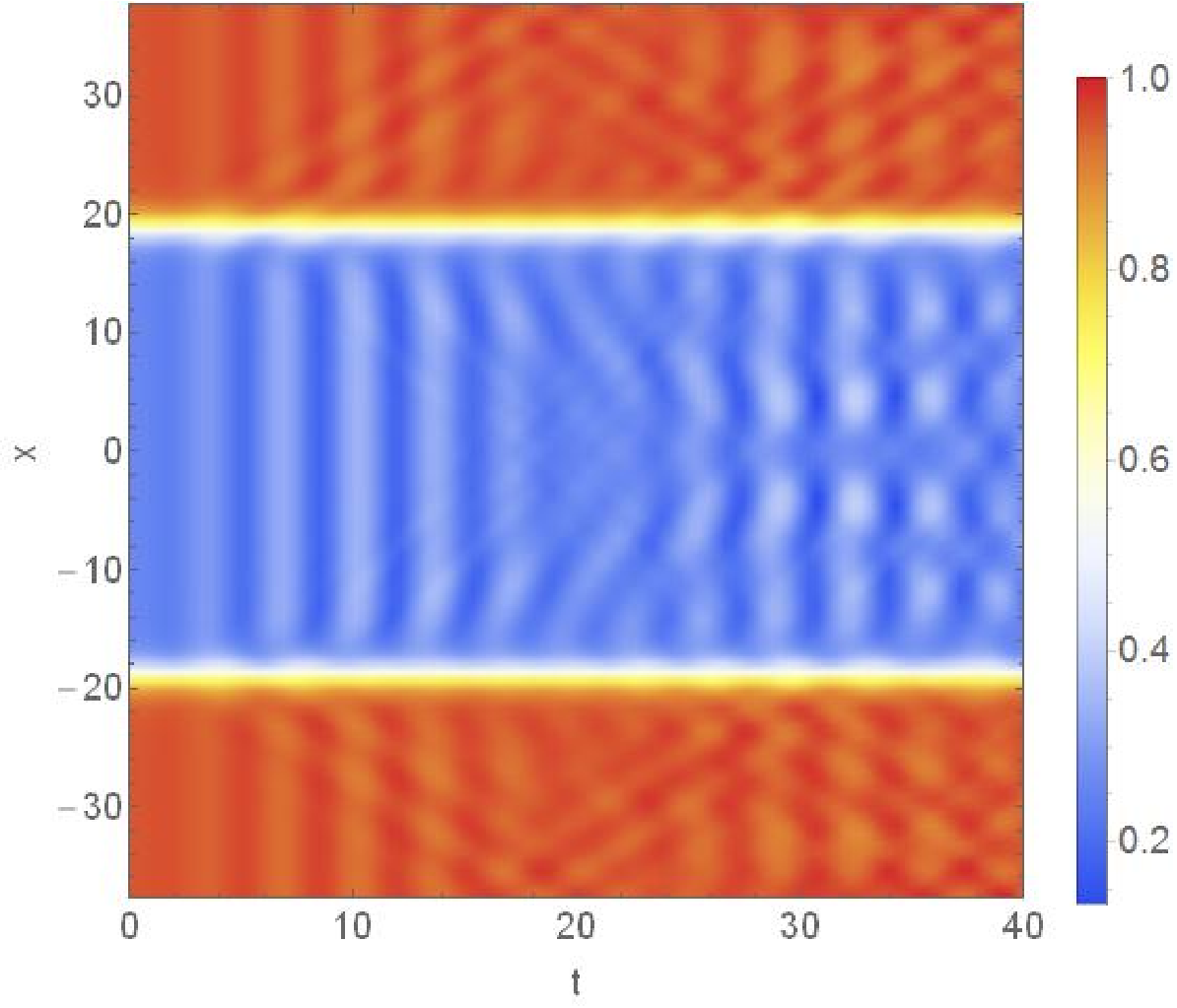}}
\caption{a) Oscillations of the atom numbers $N_{\pm }$, defined
according to Eq. (\protect\ref{Npm}), on one side of the DW
interface. The oscillations are initiated by the sudden jump
(quench) of the
cross-repulsion coefficient from $g=3$ to $g=3.3$, at fixed $\protect\kappa %
=0.5$ with input~(\protect\ref{exact1}). The simulations reveal the
eigenfrequency of the inner oscillations $\protect\omega _{0}=2$. b)
Oscillations of the binary condensate driven by NM, in the form of Eq. (%
\protect\ref{gt}) with $g=3$, $\protect\varepsilon =0.1$ and $\protect\omega %
=\protect\omega _{0}=2$. The driven oscillations display beatings,
which is the signature of the nonlinear resonance. c) The
application of the NM gives
rise to density waves, which are displayed here for component $\protect\psi %
_{+}(x,t)$.} \label{fig6}
\end{figure}

Varying the strength of the cross-repulsion $g(t)$ alters the width
of the DW's interface, with larger $g$ corresponding to a narrower
DW, and vice versa. The periodic variation of $g(t)$ leads to a
similar variation in the number of atoms in the two components,
\begin{equation}
N_{\pm }(t)=\int_{L/4}^{L/2}|\psi _{\pm }(x,t)|^{2}dx  \label{Npm}
\end{equation}%
on either side of the DW, due to the pulling or pushing the atoms
from/to the phase boundary. The effect is more pronounced, as a
resonant one, when the NM is applied at a frequency close to an
eigenfrequency of the inner oscillations of the DW. The latter one
can be identified from the results of simulations of a slightly
perturbed system. In Fig.~\ref{fig6}a) we show the oscillations of
$N_{\pm }$ corresponding to the quarter-ring on one side of the DW
[see Fig.~\ref{fig4}a) and Eq. (\ref{Npm})]. The oscillations are
initiated by the quench of $g$ from $3$ to $3.3$. Next, in Fig.
\ref{fig6}b) we apply the NM defined as per Eq. (\ref{gt}), with
$\omega =2$, which is chosen as the eigenfrequency identified in
Fig. \ref{fig6}a). The driven oscillations feature, in Fig.
\ref{fig6}b), the signature of the nonlinear resonance
\cite{sagdeev-book}, namely, periodic beatings of the oscillation
amplitude due to periodic tuning to the resonance and detuning from
it.

\subsection{The management of the MIM transition in the ring geometry}

The experimental realization of binary BECs in quasi-1D toroidal traps \cite%
{beattie2013} motivates the study of DW-anti-DW pairs in the ring
geometry. The advantage of this setting is that the absence of the
external potential makes it possible to explore the phenomenology of
the MIM transition in the \textquotedblleft pure form". DW-anti-DW
complexes in the ring geometry were also considered for binary BECs
with cubic-quintic nonlinearities and for Tonks--Girardeau binary
gases in \cite{filatrella2014}.

\begin{figure}[htbp]
\centerline{ $a)$ \hspace{5cm} $b) \hspace{6cm} c)$}
\centerline{\includegraphics[width=6cm,height=5cm]{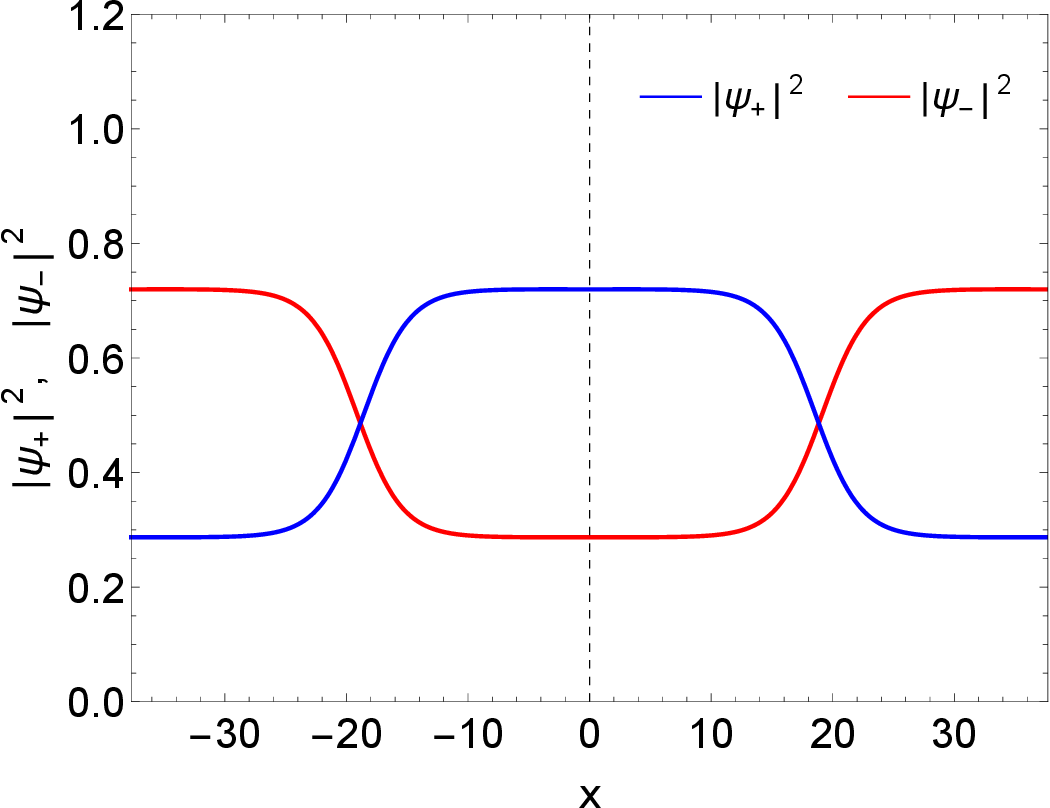}
            \includegraphics[width=6cm,height=5cm]{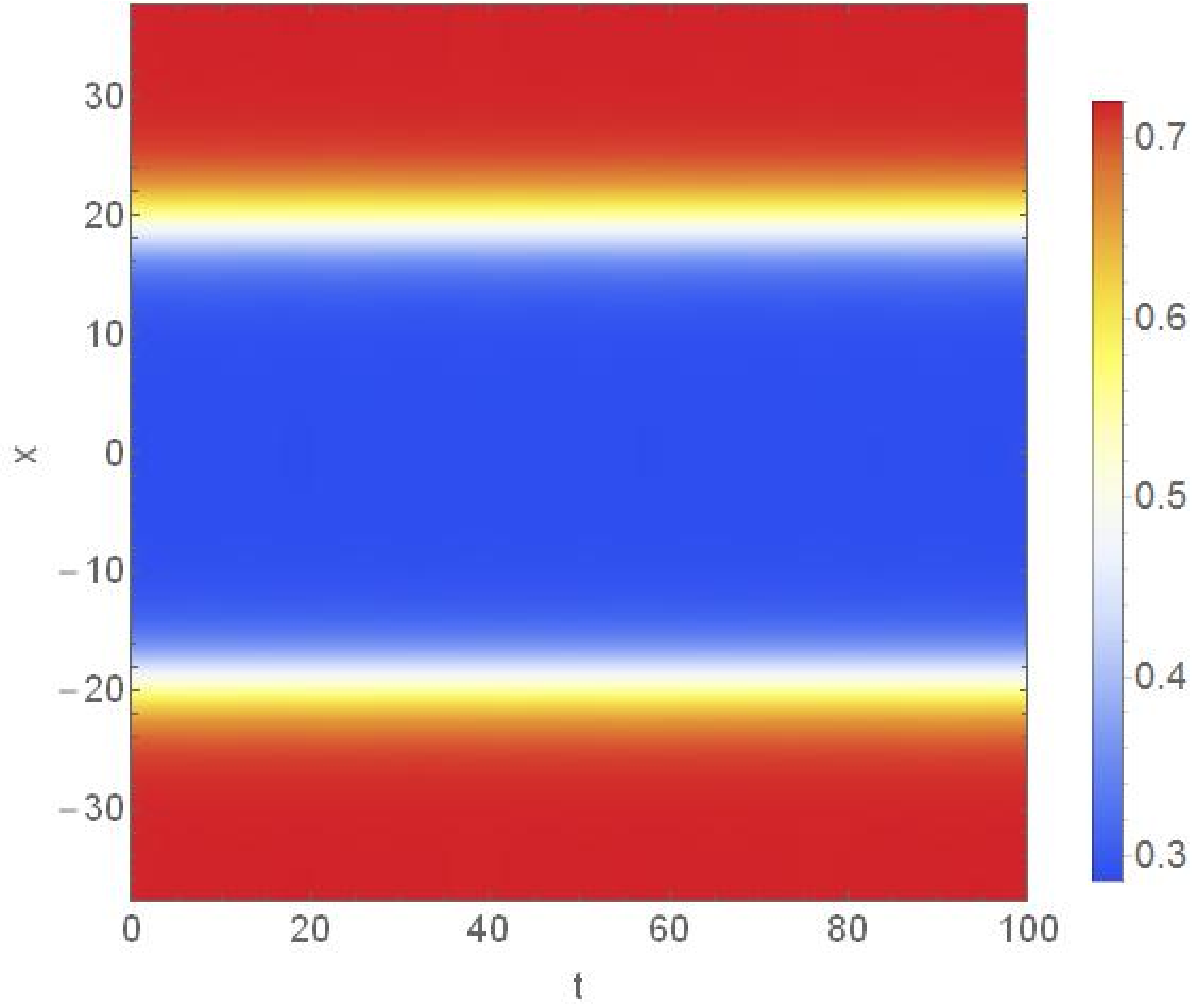}
            \includegraphics[width=6cm,height=5cm]{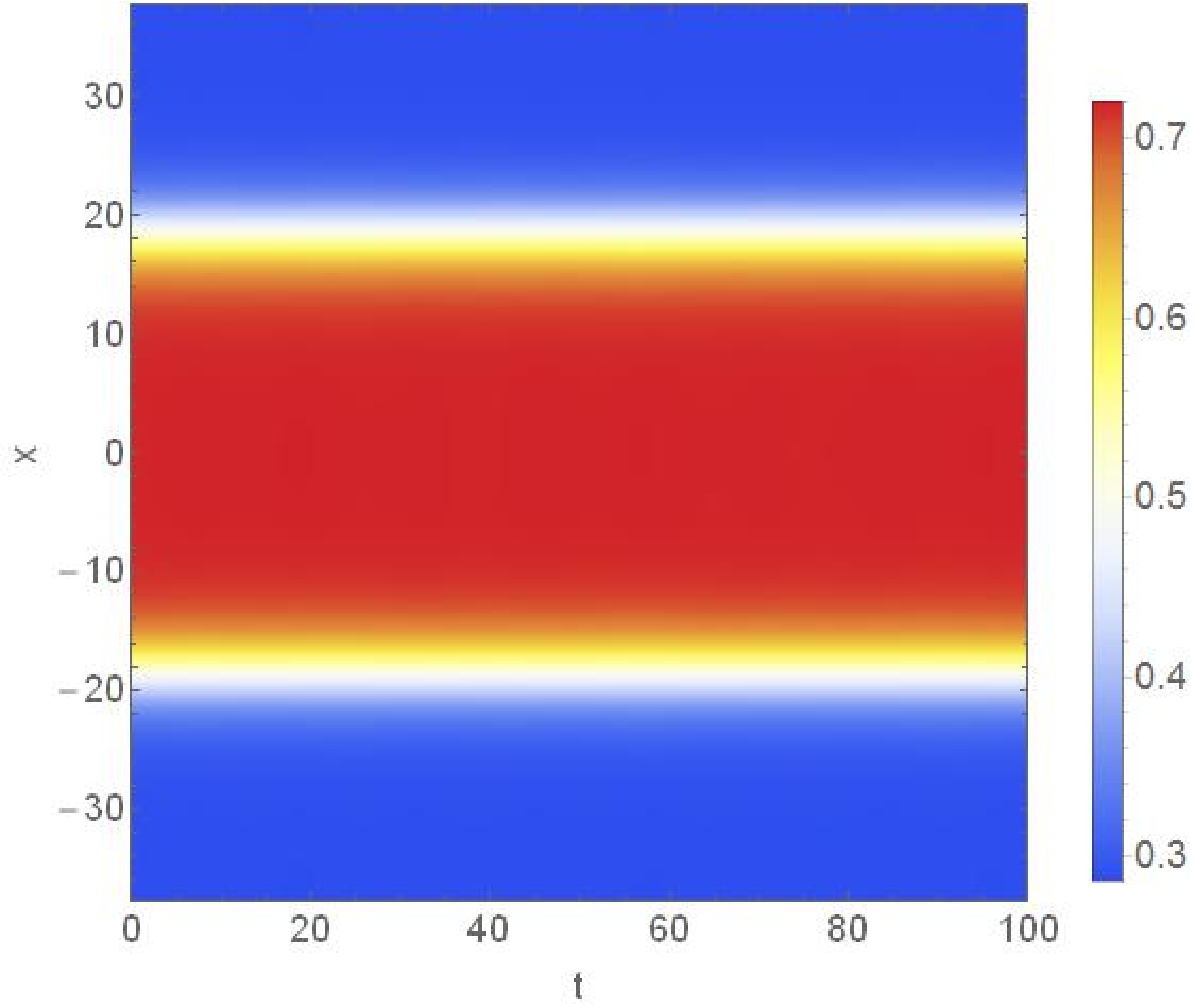}}
\caption{a) The initial density profile for simulations of the
evolution of the binary BEC in the ring-shaped setting is taken as
the numerically constructed ground state. The stable evolution of
the system is displayed by
means of the density plots $|\protect\psi _{+}(x,t)|^{2}$ in b) and $|%
\protect\psi _{-}(x,t)|^{2}$ in c). The parameter values are $g=2.1$, $%
\protect\kappa =0.5$ , and the ring's perimeter is
$L=24\,\protect\pi $.} \label{fig7}
\end{figure}

Here we address the effect of stronger NM, when the cross-repulsion
coefficient, $g(t)$, periodically passes the critical value, $g_{\mathrm{MIM}%
}$. To this end, we simulated the system (\ref{gpe1})-(\ref{gpe2}),
with the initial conditions corresponding to the numerically found
ground states of the system with constant $g=g_{0}$. This ground
state was, in turn, produced by means of the imaginary-time
integration method, starting from the approximate analytical
solution (\ref{narrow}). Using the ring geometry, we set the two
DWs, as said above, at points (\ref{L}), with $L=24\,\pi $. First,
the initial DW profiles and their stability in the absence of NM are
displayed in Fig. \ref{fig7}.

\begin{figure}[htbp]
\centerline{ $a)$ \hspace{5cm} $b) \hspace{6cm} c)$}
\centerline{\includegraphics[width=6cm,height=5cm]{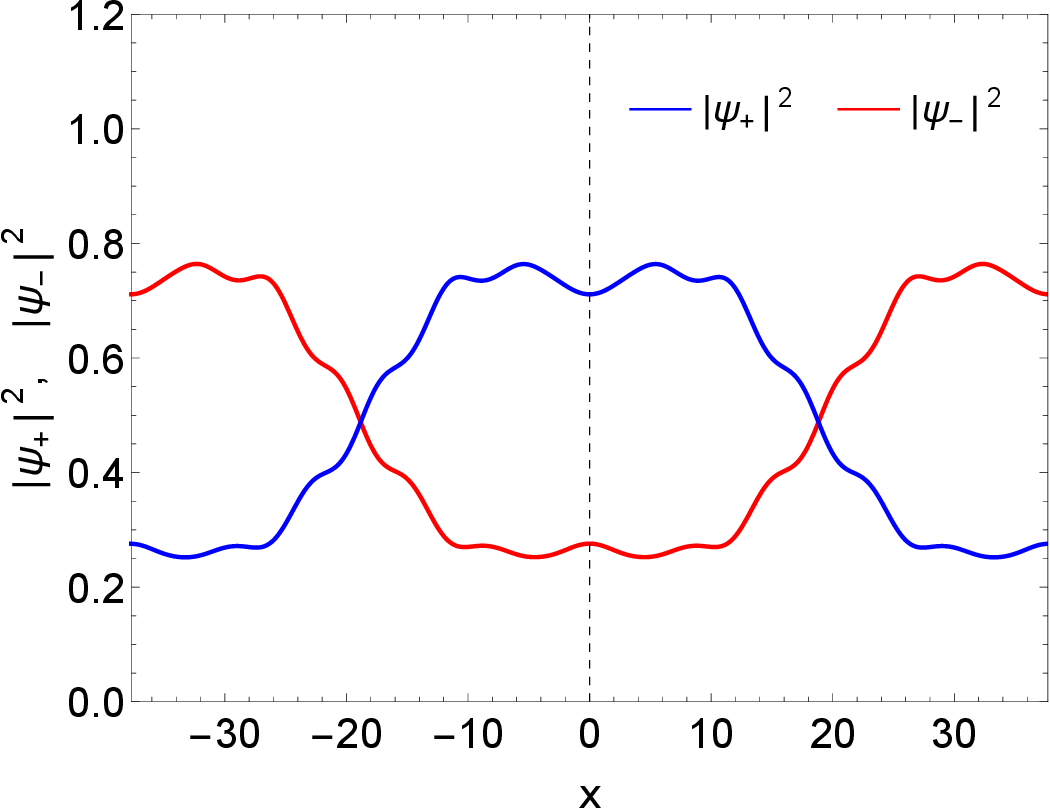}
            \includegraphics[width=6cm,height=5cm]{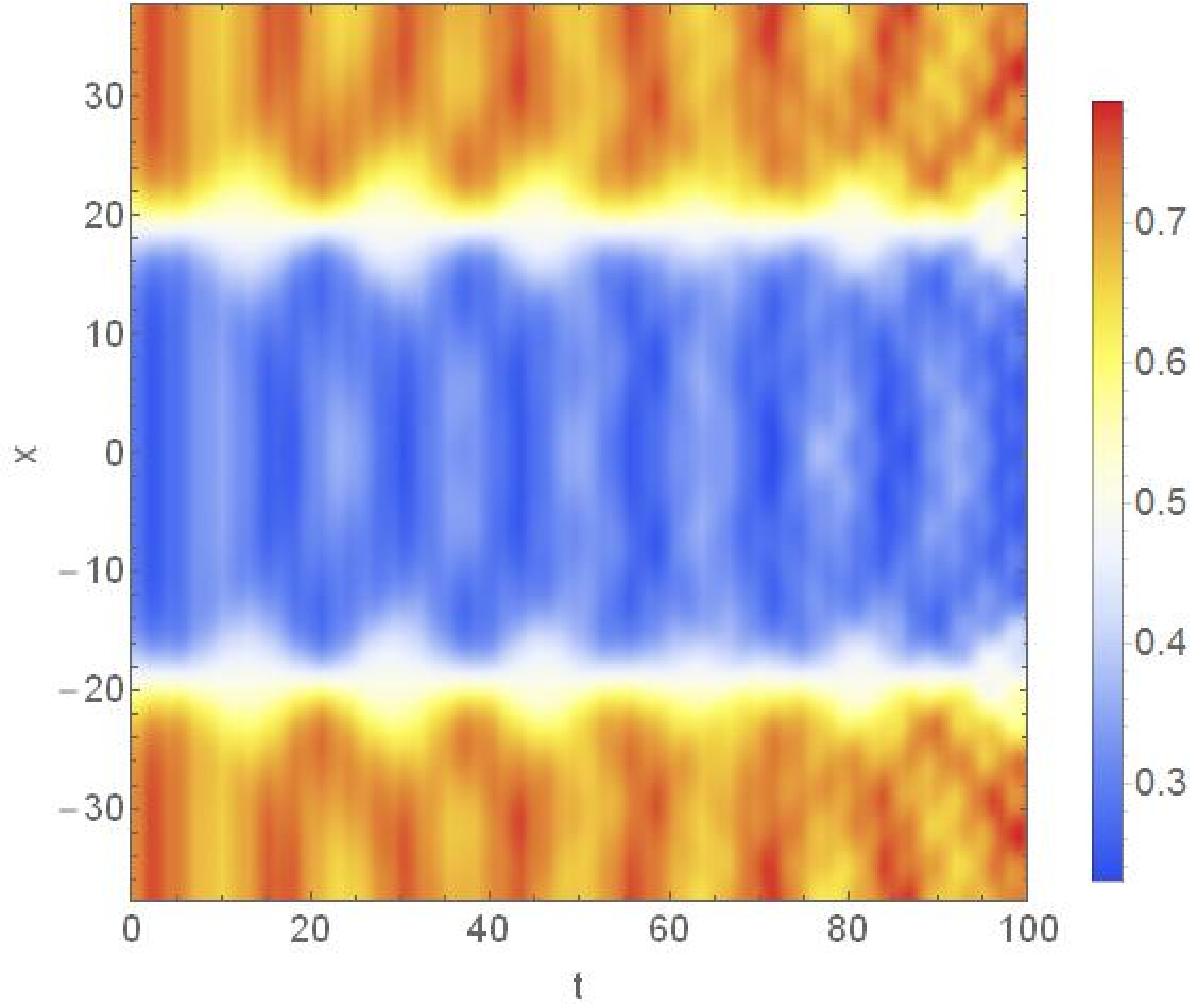}
            \includegraphics[width=6cm,height=5cm]{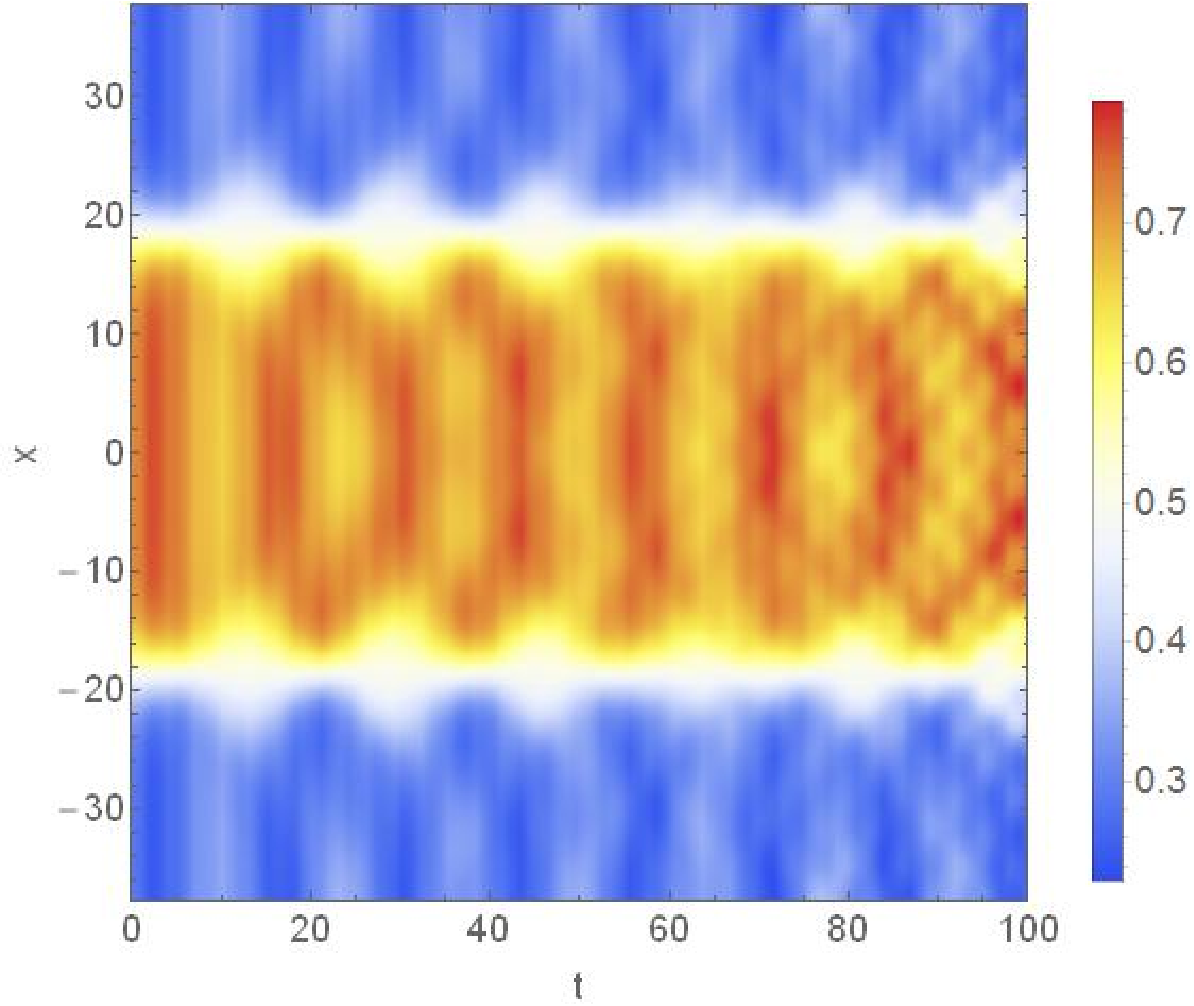}}
\centerline{ $d)$ \hspace{5cm} $e) \hspace{6cm} f)$}
\centerline{\includegraphics[width=6cm,height=5cm]{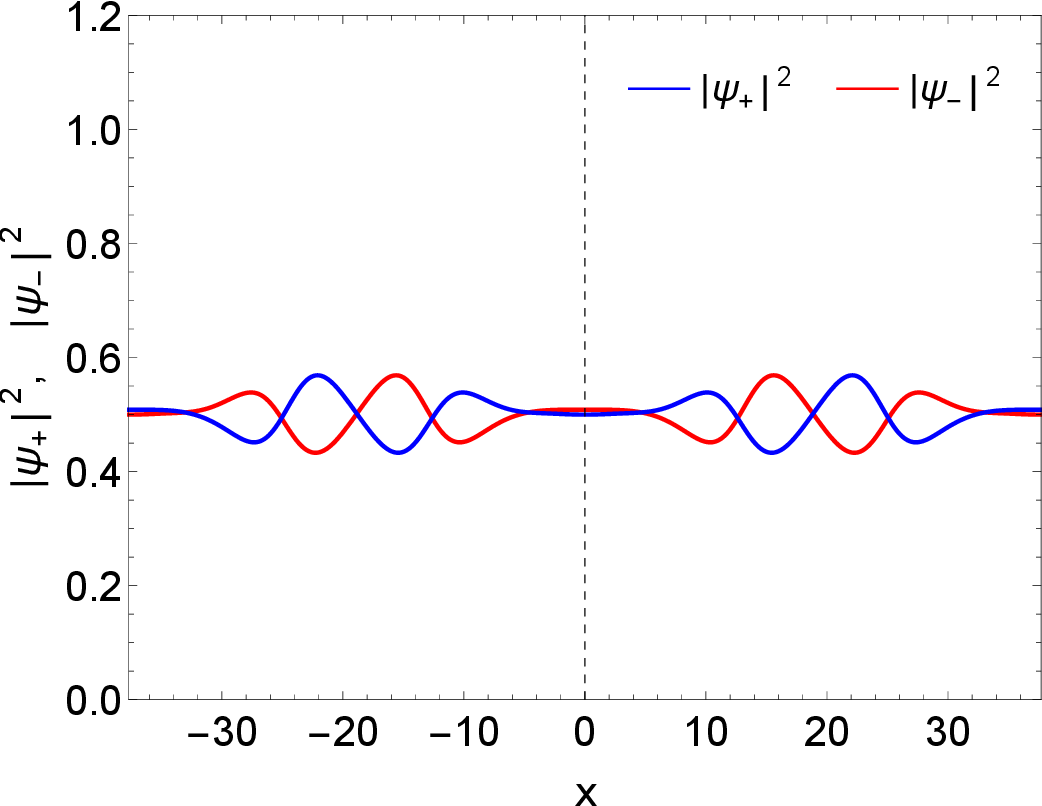}
            \includegraphics[width=6cm,height=5cm]{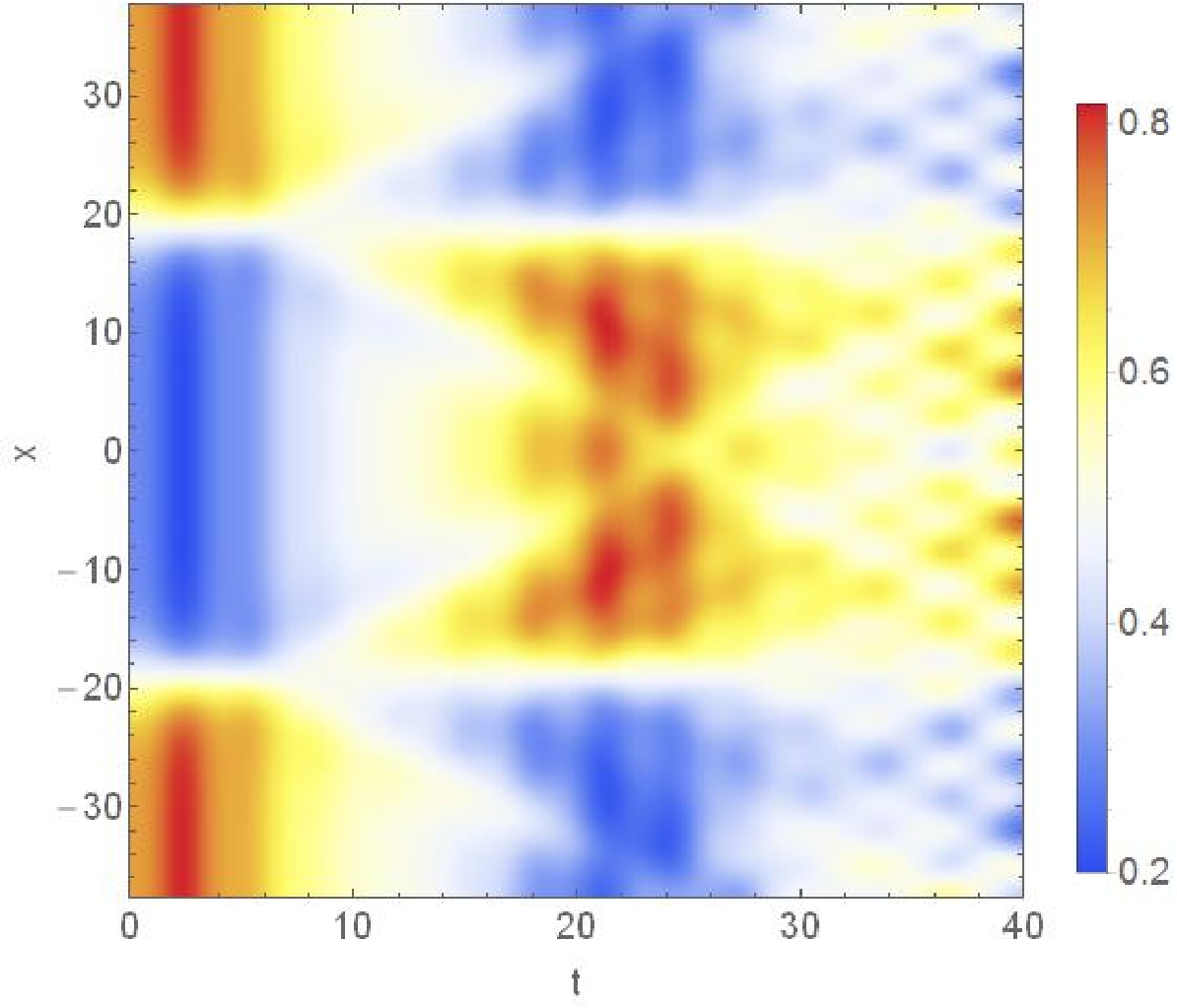}
            \includegraphics[width=6cm,height=5cm]{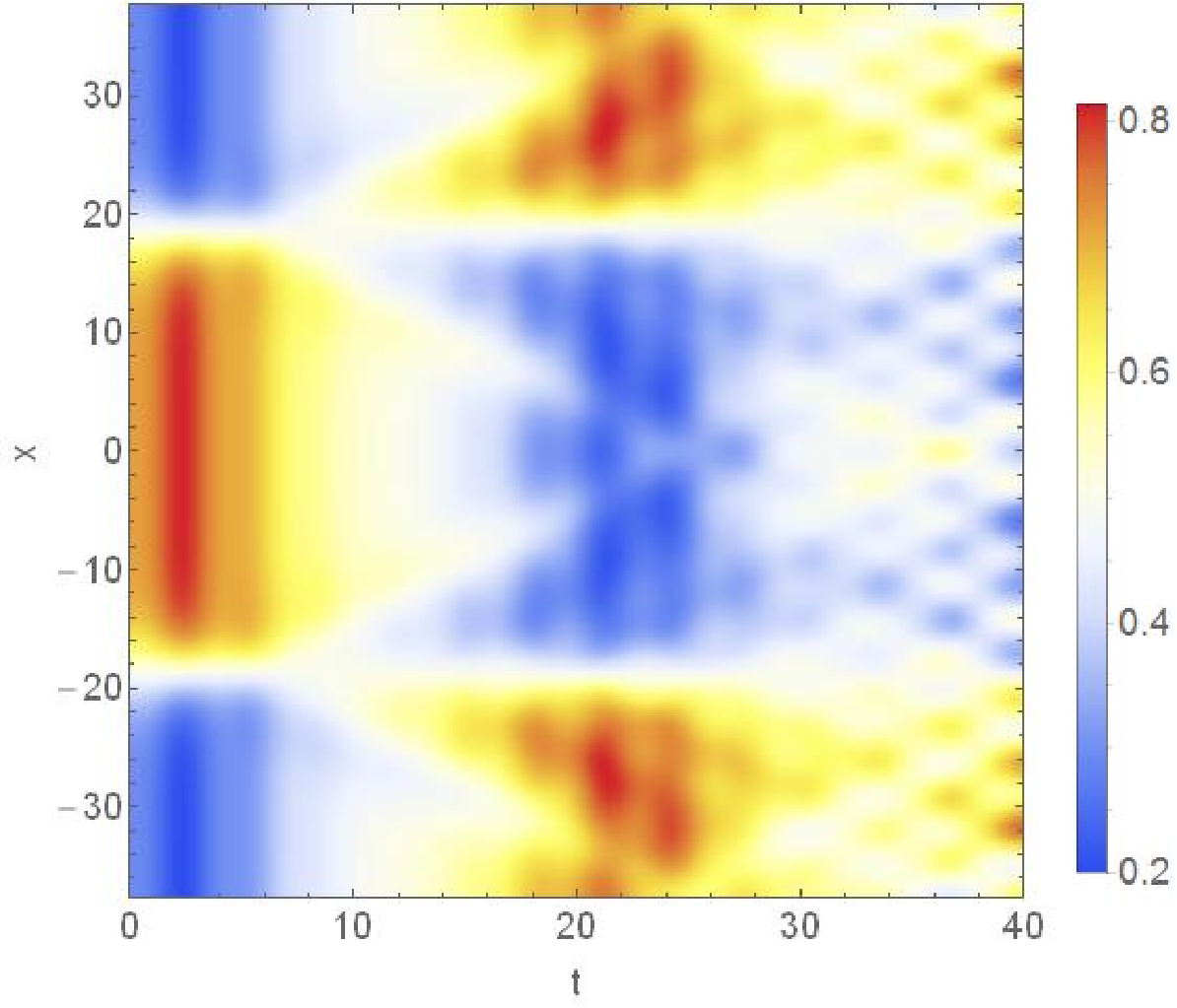}}
\caption{The density profiles of the binary condensate evolving
under the action of NM (\protect\ref{gt}), with $\protect\omega $
set equal to the
above-mentioned internal eigenfrequency $\protect\omega _{0}=2,$ and $%
g_{0}=2.1$, $\protect\kappa =0.5$. Upper panels: Moderately strong NM with $%
\protect\varepsilon =0.2$ leads to $g(t)$ periodically passing the
critical value of the MIN transition, $g_{\mathrm{MIM}}=2$. In this
case, the NM generates small-amplitude density waves, preserving the
DW structure. a) The component densities at final time $t=100$. The
spatiotemporal evolution of the density profiles of the components
$|\protect\psi _{+}(x,t)|^{2}$, b), and $|\protect\psi
_{-}(x,t)|^{2}$, c), shows their stability in the course of the
evolution. Lower panels: Stronger NM with $\protect\varepsilon =0.4$
gives rise to the transition to miscibility (hence loss of the DW
structure) around $t\simeq 10$. d) The component densities profiles
just after the onset of the transition, at $t=12$. e,f) The
evolution of the component density patterns, $|\protect\psi _{\pm
}(x,t)|^{2}$.} \label{fig8}
\end{figure}

Next, we apply the NM, pursuant to Eq. (\ref{gt}), which drives
periodic
passage of $g(t)$ through the critical value for the MIM transition, $g_{%
\mathrm{MIM}}=2$, for $\kappa =0.5$ [see Eq. (\ref{kappa})] and
monitor the response of the DW states at different amplitudes
$\varepsilon $ of the periodic modulation. Figure \ref{fig8} shows
the outcomes demonstrating stable and unstable NM regimes,
respectively. In the former case, the binary condensate remains
immiscible, while in the latter case the stronger modulation induces
a transition to miscibility, hence the DW structure is lost.

Finally, the phase diagram for the MIM transition has been produced
in the
parameter plane$\left( \varepsilon ,\omega \right) $, as shown in Fig. \ref%
{fig9}. Naturally, the increase of $\varepsilon $ favors the
transition to the miscibility, while the increase of $\omega $
attenuates the effect of the time-periodic modulations and thus
helps to keep the system in the immiscible phase.
\begin{figure}[htbp]
\centerline{\includegraphics[width=8cm,height=6cm]{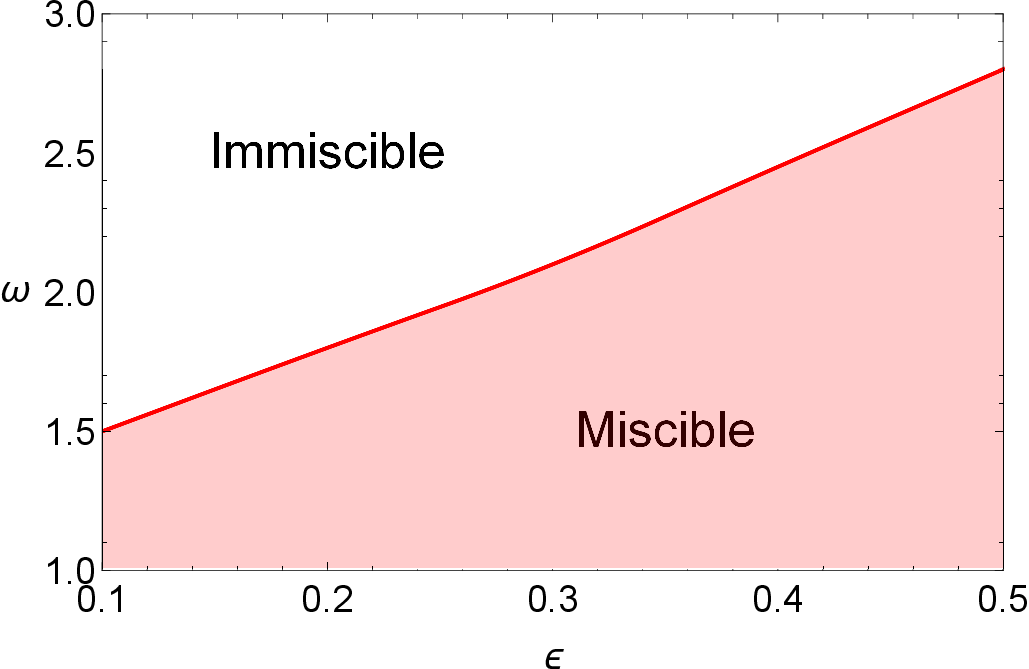}}
\caption{The phase diagram for the MIM (miscibility-immiscibility)
transition for the binary condensate under action of
NM~(\protect\ref{gt}).
The diagram is plotted in the $\left( \protect\varepsilon ,\protect\omega %
\right) $ plane. Other parameters are $g_{0}=2.1$, $\protect\kappa
=0.5$.} \label{fig9}
\end{figure}

\section{Conclusions}

In this work, we have obtained approximate analytical and numerical
solutions for DWs (domain walls) in the binary BEC with self- and
cross-repulsive nonlinearity; the system may also include the linear
RC (Rabi coupling) between the components. The DW solutions,
produced by the approximate method, based on the variational
principle, and the respective prediction of the MIM
(miscibility-immiscibility) transition, are in excellent agreement
with their numerical counterparts. All the DW states are found to be
stable. The approximate analytical solutions were also used as
efficient inputs for producing the numerical solutions for the
system's ground states, including those close to the MIM
(miscible-immiscible) transition, by means of the imaginary-time
propagation method. In addition,
a particular exact DW solution was produced for the system including the P%
\"{o}schl-Teller potential. The exact DW solution with the repulsive
(attractive) P\"{o}schl-Teller potential is found to be stable
(unstable). The numerical results were chiefly obtained for the
ring-shaped system, which carries a pair of DW and anti-DW, located
at diametrically opposite positions. A crude analytical estimate was
also reported for the shift of the MIM transition towards stronger
cross-repulsion under the pressure of the trapping potential. Then,
we addressed the system under the action of the miscibility
management, imposed by the time-periodic modulation of the
cross-repulsion strength, $g(t)$ (alias the nonlinearity management,
NM). In the case of weak NM, when $g(t)$ does not pass through the
MIM-transition point, the system demonstrates excitation of a
nonlinear resonance in the driven oscillations of the DW. Stronger
NM, which leads to periodic passage of $g(t)$ across the
MIM-transition point, may lead to a regime in which the stable DW
survives. Eventually, strong NM pulls the system into the miscible
state, and thus destroys DWs.

As an extension of the work, it may be interesting to include the
linear spin-orbit coupling and Zeeman splitting between the two
components of the binary condensate \cite{Zhai}. A promising
possibility is to consider the miscibility management in
two-dimensional systems.

\section*{Acknowledgments}

We thank Prof. Hidetsugu Sakaguchi for a valuable discussion. B.B.B.
acknowledges support from the State Budget of the Republic of
Uzbekistan (Grant No. 2025 year award) and the Innovative
Development Agency (Grant No. ALM-2023-1006-2528). The work of
B.A.M. was supported, in part, by the Israel Science Foundation
through grant No. 1695/22.


\begin{thebibliography}{99}

\bibitem{Mineev} V. P. Mineev, The theory of the solution of two near-ideal
Bose gases, Zh. Eksp. Teor. Fiz. 67, 263-272 (1974) [English
translation: Sov. Phys. -- JETP 40, 132-136 (1974)].

\bibitem{Timmermans} E. Timmermans, Phase Separation of Bose-Einstein
Condensates, Phys. Rev. Lett. 81, 5718-5721 (1998).

\bibitem{pethick-book} C. J. Pethick and H. Smith, Bose-Einstein
condensation in dilute gases (Cambridge University Press, Cambridge,
2002).

\bibitem{merhasin2005} M. I. Merhasin, B. A. Malomed, and R. Driben,
Transition to miscibility in a binary Bose-Einstein condensate
induced by linear coupling, J. Phys. B: At. Mol. Opt. Phys. 38,
877-892 (2005).

\bibitem{wang2016} F. Wang, X. Li, D. Xiong and D. Wang, A double species $%
^{23}$Na and $^{87}$Rb Bose-Einstein condensate with tunable
miscibility via an interspecies Feshbach resonance, J. Phys. B: At.
Mol. Opt. Phys. 49, 015302 (2016).

\bibitem{FR} C. Chin, R. Grimm, P. Julienne, and E. Tiesinga, Feshbach
resonances in ultracold gases, Rev. Mod. Phys. 82, 1225 (2010).

\bibitem{Inguscio} G. Roati, M. Zaccanti, C. D'Errico, J. Catani, M.
Modugno, A. Simoni, M. Inguscio, and G. Modugno, $^{39}$K
Bose-Einstein Condensate with Tunable Interactions, Phys. Rev. Lett.
99, 010403 (2007).

\bibitem{QD-ICFO} C. Cabrera, L. Tanzi, J. Sanz, B. Naylor, P. Thomas, P.
Cheiney, and L. Tarruell, Quantum liquid droplets in a mixture of
Bose-Einstein condensates, Science 359, 301-304 (2018).

\bibitem{QD-Florence} G. Semeghini, G. Ferioli, L. Masi, C. Mazzinghi, L.
Wolswijk, F. Minardi, M. Modugno, G. Modugno, M. Inguscio, and M.
Fattori, Self-bound quantum droplets of atomic mixtures in free
space, Phys. Rev. Lett. 120, 235301 (2018).

\bibitem{QD-Luca} C. D'Errico, A. Burchianti, M. Prevedelli, L. Salasnich,
F. Ancilotto, M. Modugno, F. Minardi, and C. Fort, Observation of
quantum droplets in a heteronuclear bosonic mixture, Phys. Rev. Res.
1, 033155 (2019).

\bibitem{Bakkali} B. Bakkali-Hassani, C. Maury, Y.-Q. Zou, \'{E}. Le Cerf,
R. Saint-Jalm, P. C. M. Castilho, S. Nascimbene, J. Dalibard, and J.
Beugnon, Realization of a Townes Soliton in a Two-Component Planar
Bose Gas, Phys. Rev. Lett. 127, 023603 (2021).

\bibitem{Bakkali2} B. Bakkali-Hassani, C. Maury, S. Stringari, S.
Nascimbene, J. Dalibard, and J. Beugnon, The cross-over from Townes
solitons to droplets in a 2D Bose mixture, New J. Phys. 25, 013007
(2023).

\bibitem{grain1} G. S. Rohrer, Grain boundary energy anisotropy: a review,
J. Materials Science, 46, 5881-5895 (2001).

\bibitem{grain2} H. Lim, M. G. Lee, and R. H. Wagoner, Simulation of
polycrystal deformation with grain and grain boundary effects, Int.
J. Plasticity 27, 1328-1354 (2011).

\bibitem{grain4} P. Rudolph, Dislocation patterning and bunching in crystals
and epitaxial layers - a review, Cryst. Res. Tech. 52, 1600171
(2017).

\bibitem{grain3} W. Yao, B. Wu, Y. Liu, Growth and grain boundaries in 2D
materials, ACS Nano 14, 9320-9346 (2020).

\bibitem{grain5} U. Atxitia, D. Hinzke, and U. Nowak, Fundamentals and
applications of the Landau-Lifshitz-Bloch equation, J. Phys. D:
Appl. Phys. 50, 033003 (2017).

\bibitem{grain6} E. G. Galkina and B. A. Ivanov. Dynamic solitons in
antiferromagnets, Low Temp. Phys. 44, 618-633 (2018).

\bibitem{KZ} S. Casado, W. Gonzalez-Vinas, H. Mancini, Testing the
Kibble-Zurek mechanism in Rayleigh-B\'{e}nard convection, Phys. Rev.
E 74, 047101 (2006).

\bibitem{KZ-Laroze} M. A. Miranda, D. Laroze, W. Gonzalez-Vinas, The
Kibble-Zurek mechanism in a subcritical bifurcation, J. Phys. Cond.
Matt. 25, 404208 (2013).

\bibitem{Cross1982} M. C. Cross, Ingredients of a theory of convective
textures close to onset, Phys. Rev. A 25, 1065-1076 (1982).

\bibitem{Manneville} P. Manneville, Y. Pomeau, A grain-boundary in cellular
structures near the onset of convection, Phil. Mag. A 48, 607-621
(1983).

\bibitem{Steinberg} V. Steinberg, G. Ahlers, D. S. Cannell, Pattern
formation and wave-number selection by Rayleigh-B\'{e}nard
convection in a cylindrical container, Physica Scripta 32, 534-547
(1985).

\bibitem{Trib-DW} B. A. Malomed, A. A. Nepomnyashchy, M. I. Tribelsky,
Domain boundaries in convection patterns, Phys. Rev. A 42, 7244-7263
(1990).

\bibitem{Scheel} M. Haragus, A. Scheel, Grain boundaries in the
Swift-Hohenberg equation, Eur. J. Appl. Math. 23, 737-759 (2012).

\bibitem{Iooss} M. Haragus, G. Iooss, Bifurcation of symmetric domain walls
for the B\'{e}nard-Rayleigh convection problem, Arch. Rational Mech.
Anal. 239, 733-781 (2021).

\bibitem{filatrella2014} G. Filatrella, B. A. Malomed, and M. Salerno,
Domain walls and bubble droplets in immiscible binary Bose gases,
Phys. Rev. A 90, 043629 (2014).

\bibitem{nicklas2011} E. Nicklas, H. Strobel, T. Zibold, C. Gross, B. A.
Malomed, P. G. Kevrekidis, and M. K. Oberthaler, Rabi flopping
induces spatial demixing dynamics, Phys. Rev. Lett. 107, 193001
(2011).

\bibitem{malomed2004} B. A. Malomed, H. E. Nistazakis, D. J. Frantzeskakis,
and P. G. Kevrekidis, Static and rotating domain-wall crosses in
Bose-Einstein condensates, Phys. Rev. A 70, 043616 (2004).

\bibitem{Peli} S. Alama, L. Bronsard, A. Contreras, and D. E. Pelinovsky,
Domain Walls in the Coupled Gross-Pitaevskii Equations, Arch.
Rational Mech. Anal. 215, 579--610 (2015).

\bibitem{malomed1994} B. A. Malomed, Optical domain walls, Phys. Rev. E 50,
1565 (1994).

\bibitem{malomed2021} B. A. Malomed, New findings for the old problem: Exact
solutions for domain walls in coupled real Ginzburg-Landau
equations, Phys. Lett. A 422, 127802 (2022).

\bibitem{Fluegge} S. Fl\"{u}gge, Practical Quantum Mechanics (Springer-Verlag, Berlin,
1998).

\bibitem{laser-beam} A. L. Marchant, T. P. Billam, T. P. Wiles, M. M. H. Yu,
S. A. Gardiner, and S. L. Cornish, Controlled formation and
reflection of a bright solitary matter-wave, Nature Comm. 4, 1865
(2013).

\bibitem{Campbell} D. K. Campbell, M. Peyrard, and P. Sodano, Kink-antikink
interactions in the double sine-Gordon equation, Physica D 19,
165-205 (1986).

\bibitem{CSF} M. Wang and X. Li, Exact solutions to the double Sine-Gordon
equation, Chaos, Solitons \& Fractals 27, 477-486 (2006).

\bibitem{chiofalo2000} M. L. Chiofalo, S. Succi, M. P. Tosi, Ground state of
trapped interacting Bose-Einstein condensates by an explicit
imaginary-time algorithm, Phys. Rev. E 62, 7438 (2000).

\bibitem{adhikari2002} S. K. Adhikari and P. Muruganandam, Bose-Einstein
condensation dynamics from the numerical solution of the
Gross-Pitaevskii equation, J. Phys. B: At. Mol. Opt. Phys. 35, 2831
(2002).

\bibitem{beattie2013} S. Beattie, S. Moulder, R. J. Fletcher, Z. Hadzibabic,
Persistent currents in spinor condensates, Phys. Rev. Lett. 110,
025301 (2013).

\bibitem{sagdeev-book} R. Z. Sagdeev, D. A. Usikov, G. M. Zaslavsky,
Nonlinear Physics: From the Pendulum to Turbulence and Chaos
(Harwood Academic Publishers, 1988).

\bibitem{Zhai} H, Zhai, Degenerate quantum gases with spin-orbit coupling: a
review, Rep. Prog. Phys. 78, 026001 (2015).
\end{thebibliography}
\end{document}